\documentclass[preprint]{emulateapj}
\pdfoutput=1
\usepackage{hyperref}
\hypersetup{urlcolor=blue, colorlinks=true,citecolor=blue}
\usepackage{listings}
\usepackage{amsmath}
\usepackage{multirow}

\usepackage{color}
\usepackage[normalem]{ulem}

\usepackage{subfigure}

\newcommand{\rev}[1]{{#1}}
\newcommand{\revnew}[1]{{#1}}

\graphicspath{{../figures/}}

\shorttitle{Photometric Supernova Classification With Machine Learning}
\shortauthors{Lochner et al.}

\begin{document}

\title{Photometric Supernova Classification With Machine Learning}

\author{Michelle Lochner$^1$, Jason D. 
McEwen$^2$, Hiranya V. Peiris$^1$, Ofer Lahav$^1$, Max K. 
Winter$^1$}
\thanks{Contact email: dr.michelle.lochner@gmail.com}
\affil{$^1$Department of Physics and Astronomy, University College
  London, Gower Street, London WC1E 6BT, UK}

\affil{$^2$Mullard Space Science Laboratory, University College London, Surrey RH5 6NT, UK}

\begin{abstract}
Automated photometric supernova classification has become an active area of research in recent years in light of
current and upcoming imaging surveys such as the Dark Energy Survey (DES) and the Large Synoptic Survey
Telescope, given that spectroscopic confirmation of type for all supernovae discovered will be impossible. Here,
we develop a multi-faceted classification pipeline, combining existing and new approaches. Our pipeline consists
of two stages: extracting descriptive features from the light curves and classification using a machine learning
algorithm. Our feature extraction methods vary from model-dependent techniques, namely SALT2 fits, to more
independent techniques fitting parametric models to curves, to a completely model-independent wavelet approach.
We cover a range of representative machine learning algorithms, including naive Bayes, \emph{k}-nearest neighbors,
support vector machines, artificial neural networks and boosted decision trees (BDTs). We test the pipeline on
simulated multi-band DES light curves from the Supernova Photometric Classification Challenge. Using the
commonly used area under the curve (AUC) of the Receiver Operating Characteristic as a metric, we find that the
SALT2 fits and the wavelet approach, with the BDTs algorithm, each achieves an AUC of 0.98, where 1 represents
perfect classification. We find that a representative training set is essential for good classification, whatever the
feature set or algorithm, with implications for spectroscopic follow-up. Importantly, we find that by using either the
SALT2 or the wavelet feature sets with a BDT algorithm, accurate classification is possible purely from light curve
data, without the need for any redshift information.
\end{abstract}


\section{Introduction}
Astronomy is entering an era of deep, wide-field surveys and massive datasets, which requires 
the adoption of new, automated techniques for data reduction and analysis. In the past, supernova 
datasets were 
small enough to allow spectroscopic follow-up for the majority of objects, confirming the type of 
each. Only type Ia's are currently used for cosmology, and the type is of course required for 
astrophysical modeling and studies. With the onset of surveys such as the Dark Energy Survey (DES) 
\citep{des2005,des2016} 
and the upcoming Large Synoptic Survey Telescope (LSST)
\citep{lsst2009}, only a small fraction of the dataset can be spectroscopically followed up. The 
current commonly used dataset, the Joint Light curve Analysis (JLA) \citep{betoule2014}, 
contains only 740 supernovae, 
while DES is expected to detect thousands \citep{bernstein2011} and LSST hundreds of thousands 
\citep{lsst2009} of supernovae. Thus 
alternative approaches to supernova science must be developed to leverage these largely photometric 
datasets.

Supernova cosmology is possible without strictly knowing the supernova type using, for example, 
Bayesian methods \citep{kunz2007,hlozek2012,newling2012,knights2013,rubin2015}. However, these 
techniques benefit from having a reasonable probability for the type of 
each object in the dataset, so some form of probabilistic classification is useful. Additionally, 
studies of 
core-collapse supernovae and other transients rely on good photometric classification. Further, the 
observing strategy for LSST has not yet been finalized and the effect of observing strategy on 
supernova classification has not yet been established. Here we outline a multi-faceted pipeline 
for photometric supernova classification. In future work, we will apply it to LSST simulations to 
understand the effect of observing strategy on classification.

Current photometric supernova classification techniques focus on empirically based template 
fitting \citep{sako2008,sako2014}. However, in the past few years there have been several 
innovative techniques proposed to address this problem (see \cite{kessler2010b}, references 
therein and \cite{ishida2012}).

Here, we apply machine learning to this problem, as a well-established method of automated 
classification used in many disciplines. As astronomical datasets become larger and more 
difficult to process, machine 
learning has become increasingly popular \citep{ball2009,bloom2012}. Machine learning techniques 
have been proposed as 
a solution to an earlier step in the supernova pipeline, that of classifying transients from 
images \citep{duBuisson2015,wright2015}. Machine learning is also already being employed at some 
level for photometric supernova classification in the Sloan Digital Sky Survey (SDSS) 
\citep{frieman2008,sako2008}, using the parameters from template-fitting as features 
\citep{sako2014}.

We investigate the effect of including host galaxy photometric redshift information in automated 
classification. Reliable 
redshifts are important for current classification techniques in order to reliably fit the
templates. However, for future surveys such as LSST, well-constrained, unbiased redshifts may be 
difficult to obtain and could negatively affect classification. A classification technique which is 
independent of redshift information could therefore be invaluable.

Additionally, we investigate the effect of providing the machine learning algorithms with a 
non-representative training set. In general, one would expect the spectroscopically confirmed 
training set to be biased in terms of brightness (and hence redshift) because it is far easier to 
obtain spectra for brighter objects. However, any classification technique is likely to 
underperform 
with a biased training set. In our analysis, we explore both representative and non-representative 
training sets.

In this paper, we investigate four different methods of extracting useful features from 
light curves, as well as five representative machine learning algorithms. In Sec.~\ref{sec:sims} 
we summarize the simulated dataset used. Sec.~\ref{sec:features} explains our feature extraction 
methods, varying from highly model dependent approaches to a completely model 
independent method. Section \ref{sec:machine} introduces the machine learning algorithms 
used in this work. We present our results in Sec.~\ref{sec:results} and conclude in 
Sec.~\ref{sec:conclusions}.

\section{Supernova Simulations}
\label{sec:sims}
To test and compare methods, we use existing supernova simulations from the 
Supernova Photometric Classification Challenge (SPCC) \citep{kessler2010a,kessler2010b}, which are 
simulated DES light curves built from an existing library of 
non-Ia templates and using both SALT2 \citep{guy2007} and MLCS2k2 \citep{jha2007} type Ia models. 
Simulated photometric host galaxy redshifts are available for each object and we investigate the 
effect of withholding this information from the machine learning algorithms, in order to 
see if redshift is an essential ingredient for classification. We have not tested 
whether or not the results would improve if spectroscopic redshifts were available, since for DES 
and LSST this is an unlikely scenario.

\begin{figure}
\centering
 \includegraphics[width=0.99\columnwidth]{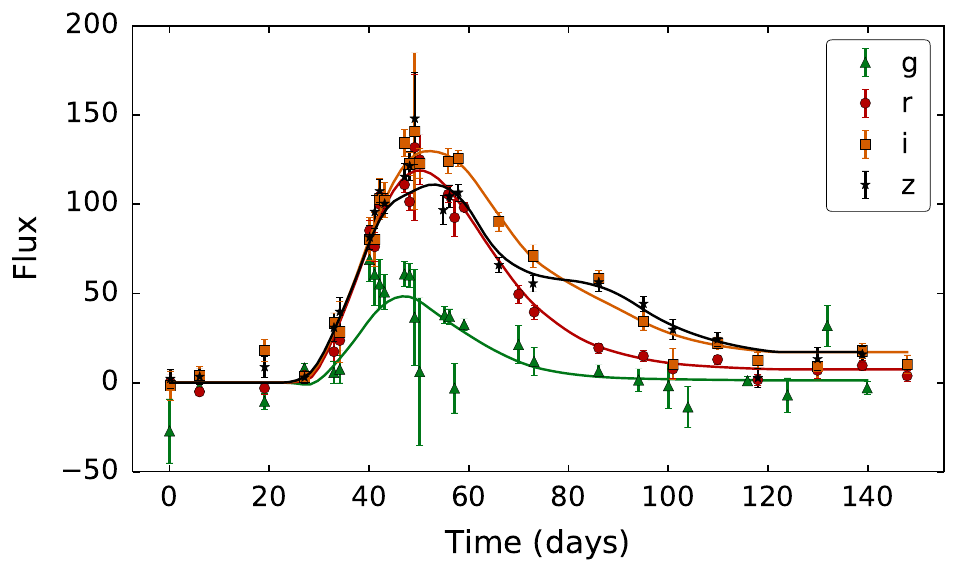}
 \caption{An example simulated DES type Ia supernova light curve (object ID 009571), at redshift 
0.42, from the Supernova Photometric Classification Challenge \citep{kessler2010a,kessler2010b}. 
The 
photometric light curve is measured in the four DES \emph{griz} filter bands, showing the rise in 
brightness after the explosion of the star and subsequent decay. The points and error bars are the 
data points, while the curves are from the best fitting SALT2 model (see Sec.~\ref{sec:features}).}
\label{fig:example_light_curve}
\end{figure}

In the original challenge, a training set of 1103 
spectroscopically confirmed objects was provided while a test set of 20216 objects was retained to 
compare the techniques under study. It is important to note that this training set is 
non-representative in brightness and hence also in redshift. This was done to emulate the way 
spectroscopic follow-up is currently performed, which prioritizes bright objects. We return to this 
point in Sec.~\ref{sec:results}. Figure \ref{fig:example_light_curve} shows an example type Ia 
light curve from the challenge data.

\section{Extracting Useful Light Curve Features}
\label{sec:features}
Feature extraction is a crucial step in the machine learning process. Raw data almost inevitably 
requires dimensionality reduction or some form of summarizing of key features in the data, and very 
few data sets can be successfully classified in a raw format. Good feature extraction techniques 
will produce a feature set of much lower dimensionality than the original data with features that 
adequately describe the original data and are well-separated between classes. Examples of commonly 
used features include principal component analysis (PCA) \citep{pearson1901,hotelling1933} 
coefficients or derived quantities such as amplitude, period, phase etc. For a general overview of 
feature 
extraction\rev{, including methods of evaluating class-separability,} see \cite{li2016}.  

We test a variety of feature sets and evaluate their performance based on the
effectiveness of subsequent classification for a variety of machine learning
algorithms. Alternative feature selection approaches agnostic of machine
learning algorithms could be considered although as these approaches evaluate the feature sets in 
isolation, they could lead to biases after a machine learning algorithm is applied. A detailed 
investigation
of alternative approaches to choosing and evaluating features is beyond the
scope of this article.

\revnew{An alternative approach to feature extraction and subsequent machine learning classification 
is deep learning \citep{schmidhuber2014,lecun2015}, which attempts to automatically learn summary 
features. Despite this 
advantage, deep learning has high computing requirements and needs a much larger training set than 
traditional algorithms. We found traditional feature extraction methods more than adequate for 
this problem. We thus consider the added flexibility and complexity of the deep learning approach, while worth exploring further, 
unnecessary for this application at present.}

We explore three different, broad approaches to feature extraction, 
ranging from a highly model-dependent method based on supernova templates to a completely 
model-independent wavelet based approach, with a parametric approach somewhere in between. 
These bracket a wide range of approaches one could consider using to extract features 
from light 
curves. We run a 
pipeline whereby we extract the four feature sets from the same data, and then run five different 
machine learning algorithms (described in Sec.~\ref{sec:machine}) on each feature set to compare 
performance. 

Before discussing the feature extraction methods we have studied, we introduce a useful 
visualization technique for understanding feature sets.

\subsection{Visualizing Feature Sets with t-distributed Stochastic Neighbor Embedding (t-SNE)}
Although feature extraction generally reduces dimensionality from raw data, the number of features 
often remains large, making it difficult to determine whether or not \rev{a particular feature 
set separates the classes well}. 
t-SNE \citep{vandermaaten2008} is a sophisticated 
technique designed to produce a low-dimensional representation of a high-dimensional space, whilst 
clustering similar features together. t-SNE works by first computing, for each pair of 
points, the probability that the two points are similar based on their Euclidean distance. The aim 
then is to find the corresponding low-dimensional values for these points that preserves the same 
probability. Stochastic neighbor embedding (SNE) determines these values by minimizing the sum of 
Kullback-Leibler divergences (a simple measure of divergence between probability distributions) over 
all data points, using a gradient descent algorithm \rev{with a stochastic element to avoid local 
minima}. t-SNE differs from SNE in the choice of exact 
probability distribution defining the similarity between 
points (using a Student-t distribution instead of a Gaussian) and in the exact choice of cost 
function to minimize (see \cite{vandermaaten2008} for details). If \rev{the classes} are 
well-separated on the t-SNE plot, one can generally expect accurate classification. 
However, the converse is not strictly true and \rev{a feature set with poor class-separation} may 
still be 
well-classified with a sophisticated machine learning algorithm. Each of the feature extraction 
methods discussed below has an accompanying t-SNE plot for visualization purposes.

\subsection{Template Fitting Approach to Classification}
\label{sec:salt2}
Historically, classification of photometric supernovae (generally used to identify candidates for 
spectroscopic follow-up) has focused on the use of supernova templates constructed from 
existing data sets of supernovae \citep{sako2008}. Light curve features such as stretch and color 
parameters have been used to classify photometric supernovae from SDSS \citep{sako2011}. We choose 
the SALT2 (Spectral Adaptive Light curve Template 2) model \citep{guy2007} of type Ia supernovae as 
it is the most commonly used.

In the SALT2 model, the flux in a given band of a type Ia light curve is given by
\begin{equation}
  \label{eq:salt2}
\begin{split}
 F(t,\lambda) = x_0 &\times [M_0(t,\lambda) + x_1M_1(t,\lambda) + ...] \\
 &\times \rm{exp}[cCL(\lambda)],
\end{split}
\end{equation}
where $t$ is the phase (time since maximum brightness in the \emph{B}-band), $\lambda$ is the rest 
frame 
wavelength, $M_0(t,\lambda)$ is average spectral sequence (using past supernova data as described 
in \citealt{guy2007}), and $M_k(t,\lambda)$ for $k>0$ are higher order components to capture 
variability in the supernovae. In practice, only $M_0(t,\lambda)$ and $M_1(t,\lambda)$ are used. 
$CL(\lambda)$ is the average color correction law. For each object, the redshift $z$ is also used 
as either a given or fitted parameter. Thus this method has 5 features: $z$, $t_0$ (the time of 
peak brightness in the \emph{B}-band), $x_0$, $x_1$ and $c$. We use the implementation of SALT2 in 
\texttt{sncosmo}\footnote{\url{http://sncosmo.readthedocs.org/en/v1.2.x/}} \citep{barbary2014} and 
obtain the best fitting parameters using 
\texttt{MultiNest}\footnote{\url{https://ccpforge.cse.rl.ac.uk/gf/project/multinest/}} 
\citep{feroz2008,feroz2009,feroz2013} with the 
\texttt{pyMultiNest}\footnote{\url{https://johannesbuchner.github.io/PyMultiNest/}} 
\citep{buchner2014} 
software. Table \ref{tab:salt2} lists the priors used on all parameters.

\begin{table}
\centering

\caption{Priors on the SALT2 model parameters.}
\label{tab:salt2}
 \begin{tabular}{cc}
 \hline
  Parameter Name & Prior\\
 \hline
  $z$ & $\mathcal{U}(0.01,1.5)$\\
  $t_0$ & $\mathcal{U}(-100,100)$\\
  $x_0$ & $\mathcal{U}(-10^{-3},10^{-3})$\\
  $x_1$ & $\mathcal{U}(-3,3)$\\
  $c$ & $\mathcal{U}(-0.5,0.5)$\\
  \hline
 \end{tabular}
 \vspace{6pt}
\end{table}

\begin{figure*}
\centering
\subfigure[SALT2 without redshift (5 features)]{
 \label{fig:tsne_salt2_no_z}
 \includegraphics[width=0.45\textwidth]{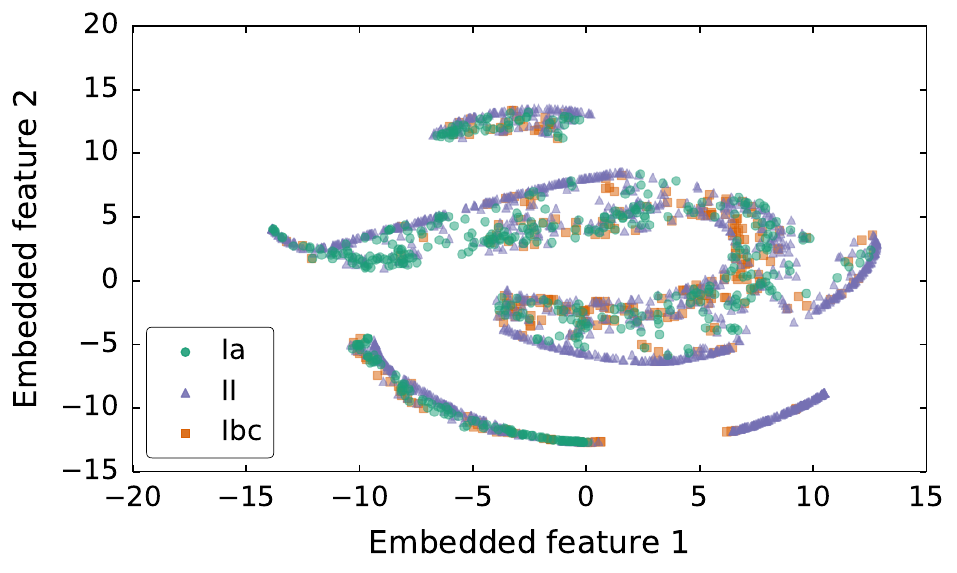}}
 \subfigure[SALT2 with redshift (5 features)]{
 \label{fig:tsne_salt2_z}
 \includegraphics[width=0.45\textwidth]{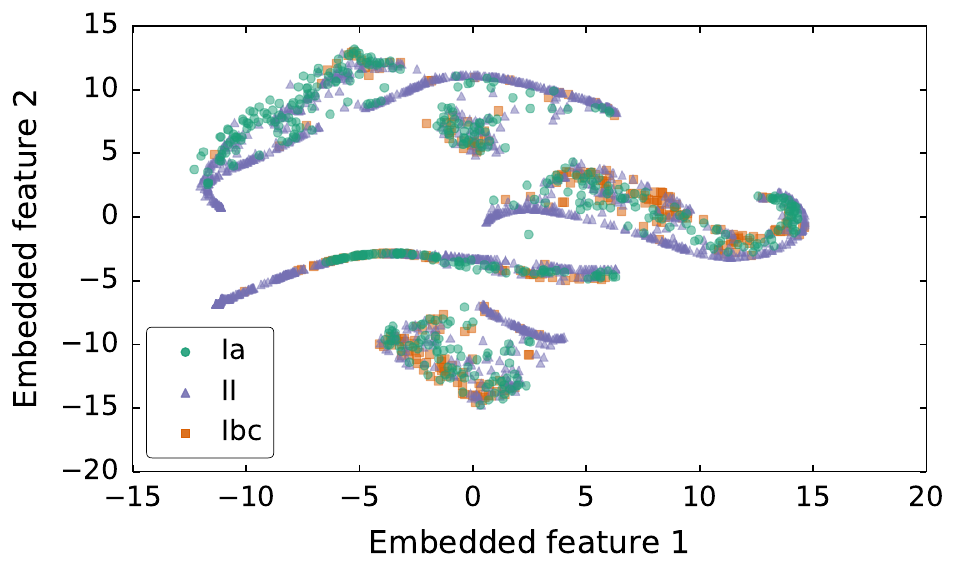}}
 
 \subfigure[Parametric Model 1 without redshift (20 features)]{
 \label{fig:tsne_newling_no_z}
 \includegraphics[width=0.45\textwidth]{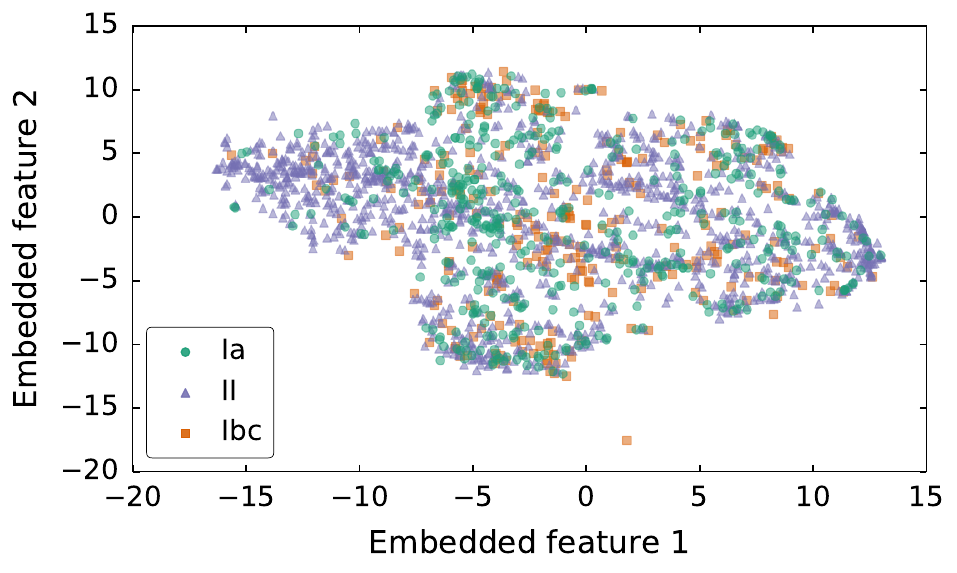}}
 \subfigure[Parametric Model 1 with redshift (21 features)]{
 \label{fig:tsne_newling_z}
 \includegraphics[width=0.45\textwidth]{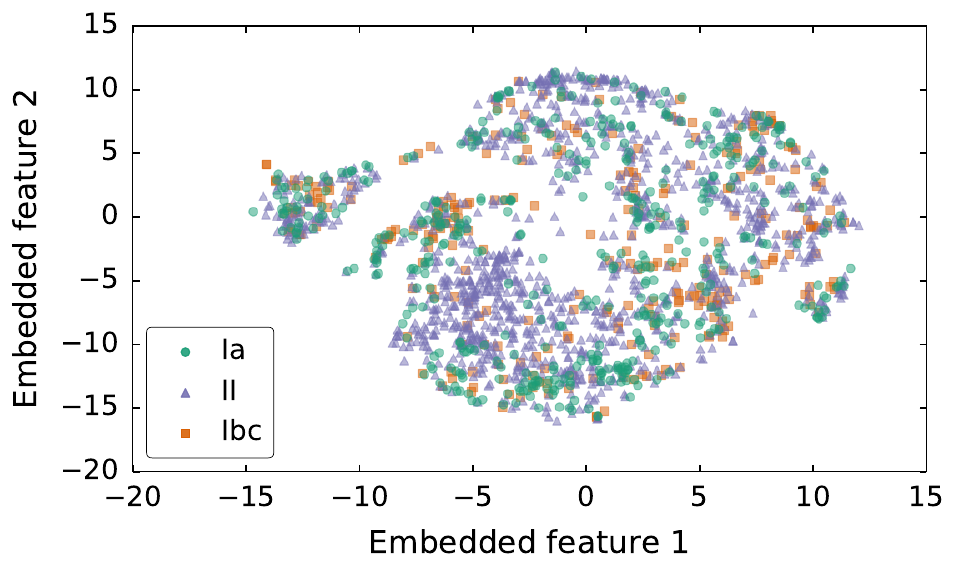}}
 
  \subfigure[Parametric Model 2 without redshift (24 features)]{
 \label{fig:tsne_karpenka_no_z}
 \includegraphics[width=0.45\textwidth]{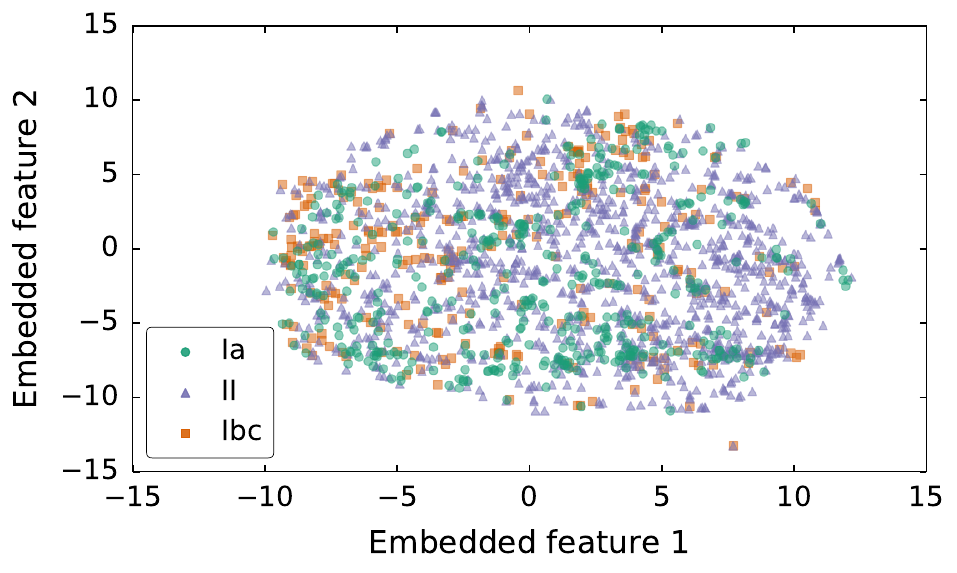}}
 \subfigure[Parametric Model 2 with redshift (25 features)]{
 \label{fig:tsne_karpenka_z}
 \includegraphics[width=0.45\textwidth]{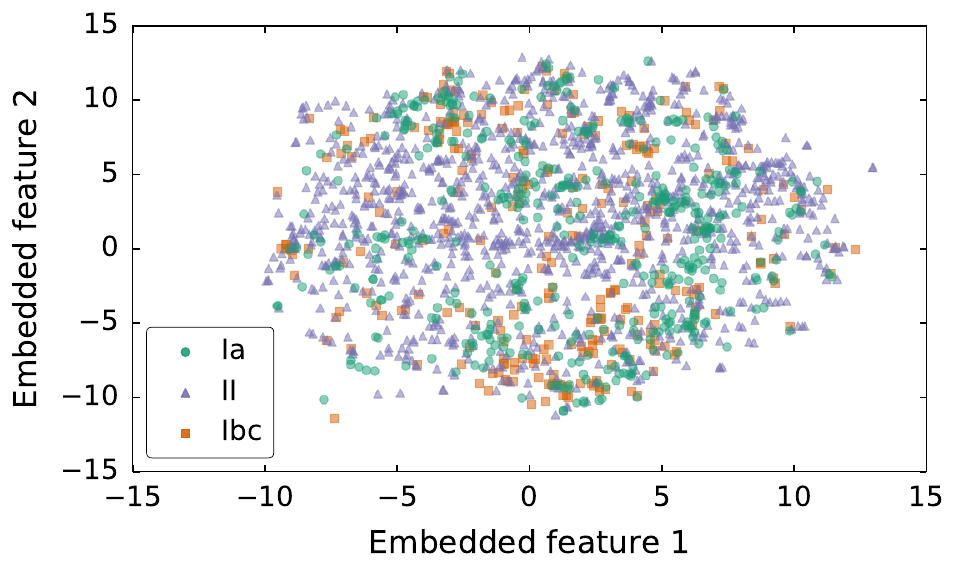}}
 
 \subfigure[Wavelets without redshift (20 features)]{
 \label{fig:tsne_wavelets_no_z}
 \includegraphics[width=0.45\textwidth]{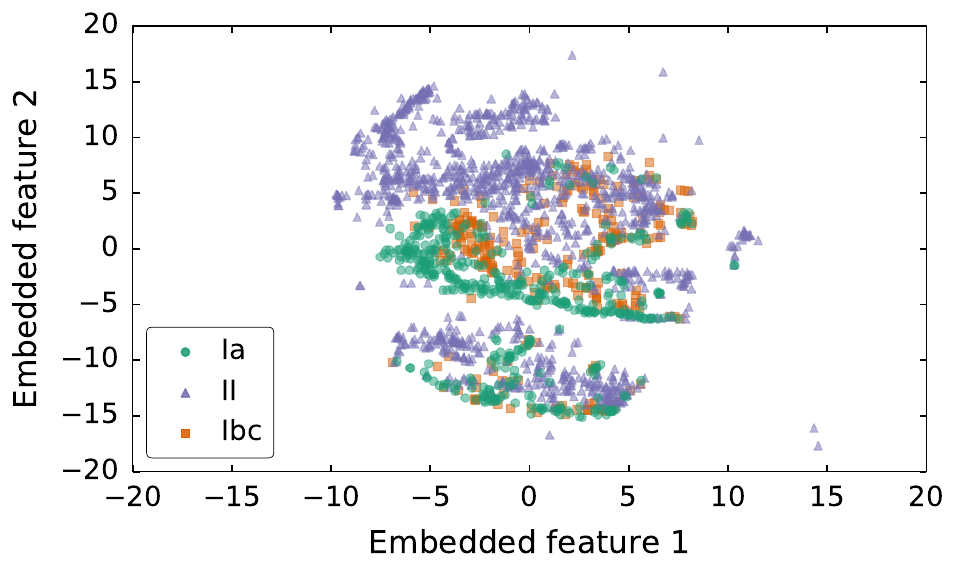}}
 \subfigure[Wavelets with redshift (21 features)]{
 \label{fig:tsne_wavelets_z}
 \includegraphics[width=0.45\textwidth]{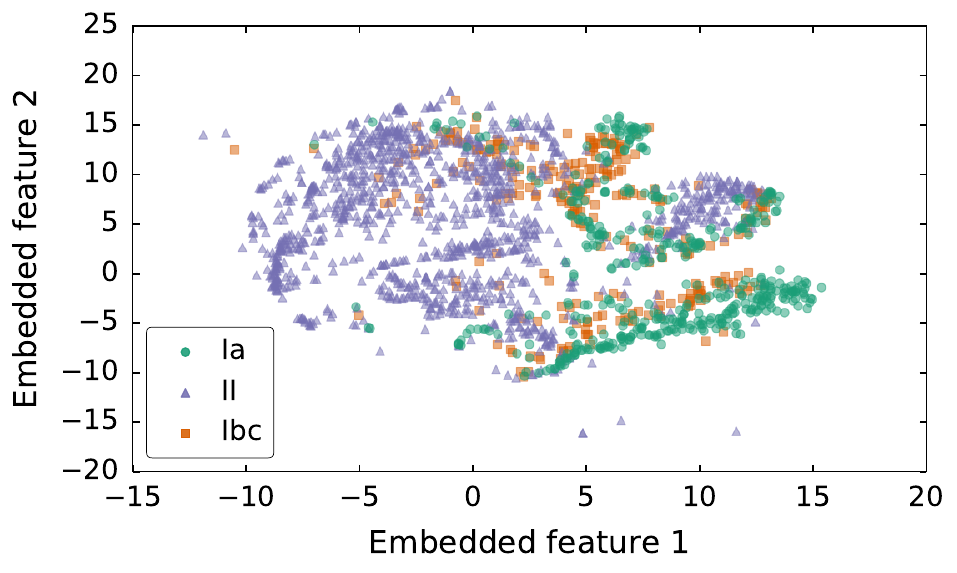}}

\label{fig:tsne}
\caption{t-SNE (t-distributed stochastic neighbor embedding) visualizations for feature sets with 
and without redshift information. This is a two-dimensional visualization of the high-dimensional 
feature space. Each of the three classes of supernovae is represented by a different color; if 
the classes are well-separated in the embedding space, we would expect these features to provide 
good classification. Note 
that all units are arbitrary and the original dimensionality is indicated in brackets. \rev{The 
reader is reminded that these plots are for visualization purposes only and the full 
classification pipeline is required to draw conclusions about the success of any feature set for 
classification.}}
\end{figure*}

Figures \ref{fig:tsne_salt2_no_z} and \ref{fig:tsne_salt2_z} show the t-SNE plots for the SALT2 
features both without and with redshift information. The t-SNE plots show a lot of structure, 
partly due to the relative low-dimensionality of the feature space, which corresponds to a highly 
constrained 2D plot. The \rev{classes} are fairly well-separated, especially when redshift is 
included, 
but it is clear that it is difficult to distinguish between type Ia and Ibc supernovae.

\subsection{General Parametric Approach to Classification}
The template-fitting approach relies on a significant amount of prior knowledge about supernovae. 
An 
alternative is to use a more general parametric description and use the fitted parameters as 
features in the machine learning pipeline. We use two different parametric models that have been 
proposed in the literature as good parameterizations of a supernova light curve and were both used 
in solutions to the SPCC. Model 1 is the 
parameterization used in \cite{newling2011}, while Model 2 is proposed in 
\cite{karpenka2013}.

We use these models because they are both proposed solutions to precisely the problem of 
supernova classification. While similar, the models have some differences, especially in the 
treatment of the tail of the light curves, the ability to describe double peaks and the 
small difference in number of parameters, which makes it interesting to compare them.

We also considered the linear piecewise model proposed in \cite{sanders2015} but found that the 
relatively large number of parameters (11 per filter) made it less robust when fitting to data, and 
consequently classification with this feature set did not typically perform well.

\subsubsection{Model 1}
The model used in \cite{newling2011} describes the flux of a supernova (in any band) by
\begin{equation}
 F(t) = A \left(\frac{t-\phi}{\sigma} \right)^k \text{exp} \left(- \frac{t-\phi}{\sigma} \right) 
k^{-k} e^k + \Psi(t),
\end{equation}
where 
\begin{equation}
 \Psi(t) = 
 \begin{cases}
  0 &-\infty < t < \infty\\
  \text{cubic spline} &\hphantom{\infty}\phi < t < \tau \\
  \psi &\hphantom{\infty}\tau < t < \infty \\
 \end{cases},
\end{equation}
and $A + \psi$ is the peak flux, $\phi$ is the starting time of the explosion, $k$ governs the 
relative rise and decay times of the light curve and $\sigma$ is the width or `stretch' of the 
light curve. The time of the peak flux, $\tau$, is given by $\tau=k\sigma + \phi$. The function 
$\Psi(t)$ is a `tail' function that ensures that the flux tends to a finite value as $t \rightarrow 
\infty$ and the cubic spline is determined to have zero derivative at $t=\phi$ and $t=\tau$. 

Each filter band is fitted individually with these five parameters 
($A$, $\phi$, $\sigma$, $k$, $\psi$), giving a total of 20 features for each object (or 21 if 
redshift is included). The parameters are all fitted with \texttt{MultiNest}, as in 
Sec.~\ref{sec:salt2}, and the best fit is taken to be the maximum posterior value for each 
parameter. 
The priors on the parameters (which are the same for all four filter bands) are given in Table 
\ref{tab:newling}.
\begin{table}[h!t]
\centering
\caption{Priors on the Model 1 parameters.}
\label{tab:newling}
 \begin{tabular}{cc}
 \hline
  Parameter Name & Prior\\
 \hline
  $\text{log}(A)$ & $\mathcal{U}(0,10)$\\
  $\phi$ & $\mathcal{U}(-60,100)$\\
  $\text{log}(\sigma)$ & $\mathcal{U}(-3,4)$\\
  $\text{log}(k)$ & $\mathcal{U}(-4,4)$\\
  $\text{log}(\psi)$ & $\mathcal{U}(-6,10)$\\
  \hline
 \end{tabular}

\end{table}

Figures \ref{fig:tsne_newling_no_z} and \ref{fig:tsne_newling_z} show the t-SNE plots for the Model 
1 feature set both without and with redshift information respectively. In both cases, the features 
do not appear to be easily separable.

\subsubsection{Model 2}
The proposed parameterization from \cite{karpenka2013} is similar to Model 1 but allows for the 
possibility of double peaks in light curves. The flux is given by

\begin{equation}
 F(t) = A\bigg[1+B(t-t_1)^2\bigg] \, 
\frac{\text{exp}[-(t-t_0)/T_{\text{fall}}]}{1+\text{exp}[-(t-t_0)/T_{\text{rise}}]},
\end{equation}
where $t$ is assumed to be the earliest measurement in the $r$-band light curve. We again fit all 
parameters using \texttt{MultiNest} and take the maximum posterior parameters. There are 6 
parameters per filter band, thus forming a feature set of 24 features (or 25 including redshift) for 
each object. We use the same priors as in \cite{karpenka2013}, given in Table \ref{tab:karpenka}.

\begin{table}[h!b]
\centering

\caption{Priors on the Model 2 parameters.}
\label{tab:karpenka}
 \begin{tabular}{cc}
 \hline
  Parameter Name & Prior\\
 \hline
  $\text{log}(A)$ & $\mathcal{U}(\text{log}(10^{-5}),\text{log}(1000))$\\
  $\text{log}(B)$ & $\mathcal{U}(\text{log}(10^{-5}),\text{log}(100))$\\
  $t_0$ & $\mathcal{U}(0,100)$\\
  $t_1$ & $\mathcal{U}(0,100)$\\
  $T_{\text{rise}}$ & $\mathcal{U}(0,100)$\\
  $T_{\text{fall}}$ & $\mathcal{U}(0,100)$\\
  \hline
 \end{tabular}

\end{table}

The t-SNE plots for the Model 2 features can be seen in Fig.~\ref{fig:tsne_karpenka_no_z} and 
Fig.~\ref{fig:tsne_karpenka_z} without and with redshift respectively. As in Model 1, the features 
do not appear to be well-separated.

\subsection{Wavelet decomposition approach to classification}
While the model-dependent approaches to feature extraction considered so far are useful for 
incorporating prior information 
about SN light curves, they are sensitive to mis-specification of light curve models.  It is quite 
possible important light curve characteristics cannot be captured efficiently by simple models, such 
as Models 1 and 2 described above.  A model-independent, non-parameteric approach to feature 
extraction is therefore required as a complementary approach that is not dependent on prior 
information.

A wavelet decomposition is a good choice for feature extraction since certain wavelet 
transforms can be 
constructed that are approximately invariant under translation and stretch.
\rev{One could then expect the wavelet coefficients of two similar supernovae to be similar, even 
if the light curves have different explosion times or redshifts.}

Recently, \cite{varughese2015} have shown that a 
wavelet-based approach to photometric supernova classification can be highly effective. Here we 
consider a substantially different method in the actual wavelet decomposition of the light curves and 
subsequent classification, but also find wavelets to be an excellent feature extraction method.

The procedure we follow to extract features using wavelets is to first interpolate the light 
curves onto the same grid of points (using Gaussian processes), then to perform a redundant wavelet 
decomposition, and lastly to use PCA to reduce the dimensionality of 
the feature set. \rev{We note that there are many alternative dimensionality reduction techniques, 
such as independent component analysis \citep{comon1994}, autoencoders 
\citep{ballard1987,kramer1991} and 
Sammon 
mapping \citep{sammon1969}, which may improve our classification results, but conclude that for this 
work PCA is more than sufficient.}

\subsubsection{Interpolation Using a Gaussian Process Regression}
In order to perform a wavelet decomposition, all light curves are interpolated onto the same uniform grid
 of points. We use Gaussian process regression since it has the advantage over (for 
example) spline 
interpolation in that it allows the straightforward inclusion of uncertainty information and 
produces a less biased estimate of interpolated values.

A Gaussian process is the generalization of a Gaussian distribution, forming a distribution for 
which each point of the input variable is associated with a normally distributed random variable 
\citep{mackay2003}. A Gaussian process is fully specified by a mean function of the input variable 
(time, in the case of a light curve) and a covariance function, which specifies the covariance 
between values of the function at pairs of points in time. Gaussian process regression then 
proceeds by determining the mean function and hyperparameters of the covariance matrix. We use the 
package \texttt{GaPP}\footnote{\url{http://www.acgc.uct.ac.za/~seikel/GAPP/main.html}} used in 
\cite{seikel2012} to perform the Gaussian process regression of each 
of the light curves.

\subsubsection{Wavelet Decomposition} 
Wavelet transforms are an invaluable tool in signal analysis due to the
ability of wavelets to localize signal content in scale and time
simultaneously, whereas real space or Fourier representations provide signal
localization, respectively, in time or frequency only (see \cite{valens1999}
for a gentle introduction to wavelets).  For the continuous wavelet transform (CWT), the
set of wavelet atoms forming the transform dictionary are scale- and
translation-invariant, i.e., a scaled or translated atom also belongs to the
dictionary, hence the CWT achieves scale- and translation-invariance \citep{mallat:2009}.  However, the discrete wavelet transform (DWT), which
permits fast transforms and exact synthesis from a finite, discrete set of
wavelet coefficients, suffers from a lack of translation-invariance due to the
critical sub-sampling performed. Approximate translation-invariance can be
achieved by eliminating sub-sampling from the DWT, leading to the redundant
\`a trous wavelet transform \citep{holschneider1989,mallat:2009}, which is also called the
stationary wavelet transform.  

We adopt the \`a trous wavelet transform, which also achieves dyadic
scale-invariance, and use the symlet family of wavelets, which are a more
symmetric version of the widely used Daubechies family of wavelets
\citep{daubechies1988}.  We use the implementation provided in the \texttt{PyWavelets}\footnote{\url{http://www.pybytes.com/pywavelets/contents.html}}
software package.  Our results were found not to be highly dependent on the
family of wavelet chosen.  Alternative wavelet constructions that achieve
better translation- and scale-invariance could be considered \citep[e.g.][]{kingsbury:2001,lo:2009,xiong:2000}
but a detailed optimization of the wavelet transform used was not
performed, highlighting the robustness of our approach.

For this work, we used 100 points on the Gaussian process curve and a
two-level wavelet transform, returning 400 (highly redundant) coefficients per
filter, or 1600 coefficients  per object.

\subsubsection{Dimensionality Reduction with PCA}
The output of the wavelet decomposition is highly redundant (which preserves translation 
invariance) and thus extremely high-dimensional. The final step in the process is to run a 
PCA to reduce dimensionality. PCA is a 
linear transform that transforms a 
set of correlated variables into linearly uncorrelated variables using singular value decomposition 
of the matrix of variables. From this, a set of eigenvectors (components) and eigenvalues are 
obtained. The components with large eigenvalues describe the majority of the variability of the 
dataset and dimensionality reduction is achieved by discarding components with negligible 
eigenvalues. After PCA, we reduce the dimensionality of the feature set from 1600 to 20 dimensions
whilst retaining 98\% of the information in the dataset (as measured by determining the fraction of 
the sum of eigenvalues of components retained to that of all components).

We show the t-SNE plots for the wavelet features without and with redshift in 
Fig.~\ref{fig:tsne_wavelets_no_z} and Fig.~\ref{fig:tsne_wavelets_z}. Here, the features seem 
encouragingly well-separated between classes, even between type Ia and Ibc in some parts of feature 
space. We would thus expect the wavelet features to perform well in classification.

\section{Machine Learning for Classification}
\label{sec:machine}
Machine learning is a powerful tool for modern astronomy \citep{bloom2012,ball2009}, becoming 
increasingly popular in the era of massive data sets. It allows the automation of tasks previously 
performed by humans and also, as in this case, allows the classification of objects that are 
seemingly inseparable to the human eye. Supervised machine learning algorithms learn a mapping 
function from training data to allow them to classify new test objects. A large variety of 
machine learning algorithms exist, and it is beyond the scope of this paper to consider 
each one. Instead, we consider five popular machine learning algorithms that are fairly 
representative of the main approaches used. All these algorithms and several 
others are comprehensively compared in \cite{caruana2006}, who find (as we do) that boosted 
decision trees and random forests often outperform other classifiers for many problems. For all 
algorithms, we use the Python package 
\texttt{scikit-learn}\footnote{\url{http://scikit-learn.org/}} 
\citep{sklearn2011}. The selected algorithms used are able to compute not just the 
classification of a particular object, but also a classification probability. \rev{This probability 
can be used to apply a cut to the set of classified objects to ensure a minimum level of 
purity in a given class.}

\subsection{Description of Machine Learning Algorithms}
The algorithms used in this work are: naive Bayes (a basic probabilistic 
classifier), \emph{k}-nearest neighbors (KNN) (a classifier based on clustering of features), a 
multi-layer 
perceptron (MLP) (the simplest form of artificial neural network; ANN), a support vector machine (SVM) (which works 
by finding a separating hyperplane between classes) and BDTs (an ensemble method 
that combines a multitude of decision trees, a weaker classifier). What follows is a brief 
overview of the five algorithms considered, in each case referring the reader 
to references for more details. 

\subsubsection{Naive Bayes}
Given a set of features for an object, ${x_1,x_2,...x_n}$, the Naive Bayes (NB) algorithm 
\citep[e.g.][]{zhang2004} states 
that 
the probability of that object belonging to class $y$ is

\begin{equation}
\label{eq:nb}
 P(y|x_1,...,x_n) \propto P(y) \displaystyle \prod_i P(x_i|y),
\end{equation}
assuming independence between features. $P(y)$ can be estimated by determining the frequency of $y$ 
in the training set. In the implementation of NB that we use, the likelihood, $P(x_i|y)$, is 
assumed to 
be Gaussian, the parameters of which are determined from the training data by maximizing the 
likelihood. The predicted class is simply determined to be the class that maximizes 
Eq.~\eqref{eq:nb}. 

The advantage of NB is that it is fast and scales very well to high dimensions. The 
disadvantage is that it assumes independence between features and that 
the features are Gaussian-distributed. One or both of these assumptions is frequently broken. 
\rev{NB has recently been used in astronomy to classify asteroids to determine targets for 
spectroscopic follow-up \citep{oszkiewicz2014}}.

\subsubsection{\emph{k}-Nearest Neighbors}
KNN \citep[see e.g.][]{altman1992} is a simple algorithm which 
classifies an 
object 
by performing a majority vote of the classes of the $k$ nearest neighbors. We use a 
Euclidean distance measure to determine the nearest neighbors, applying a weight to each neighbor 
inversely proportional to the distance.

The probability of an object belonging to class $y$ is found by summing the weights of 
all neighbors also belonging to class $y$ and dividing by the sum of the weights of all neighbors. 
The advantage of KNN is its relative simplicity and its use as a clustering, unsupervised learning 
algorithm. The disadvantages are that it is computationally intensive for large datasets, given 
that all 
pairwise distances have to be calculated; further, the results are highly dependent on the exact training 
set used, making the results fairly inconsistent especially for small training sets. \rev{An 
example of a recent application of KNN in astronomy can be found in \citet{kugler2014} where it is 
applied to spectroscopic redshift estimation.}

\subsubsection{Artificial Neural Networks}
ANNs \citep[e.g.][]{jeffrey1986} are a family of machine learning 
algorithms 
inspired by biological systems of interconnected neurons. The aim of an ANN is to learn a 
nonlinear function to map input features to output classes. For this work we use the commonly used 
multi-layer perceptron (MLP) implemented in \texttt{scikit-learn}.

Each neuron transforms the inputs of the previous layer by constructing a weighted sum of the 
features, followed by some nonlinear activation function, in this case a hyperbolic tan function. 
While in theory different topologies could be used, the MLP is constructed as a series of layers of 
neurons. In each layer, each neuron is connected to all the neurons of the previous layer, but not 
any others.

The standard topology is: $n \rightarrow h_1 \rightarrow h_2 \dots h_m \rightarrow c$, where $n$ 
is the number of input features, $h_i$ is the number of neurons in layer $i$ and $c$ is the final 
class. \rev{As is often seen in the literature}, we find it does 
not improve performance to add more than one hidden layer. 
The \texttt{scikit-learn} implementation of a MLP uses backpropagation \citep{werbos1974} to 
determine the weights.

The probability of an object belonging to a particular class is easily estimated from a MLP by 
normalizing the final activation function values so that they sum to one, thus treating them as 
probabilities. Neural networks have the advantage of being capable of learning highly nonlinear 
mappings, often allowing highly accurate classifications. The disadvantages of neural networks 
\rev{are} 
their sensitivity to tuning of hyperparameters, such as number of hidden layers and the number of 
nodes in each layer, \rev{and their tendency to overfit the data}. \rev{There are many examples of 
the use of ANNs in astronomy, \revnew{including galaxy classification} \citep{lahav1995} and their 
recent application to photometric redshift estimation \citep{sadeh2015}.}

\subsubsection{Support Vector Machines}
SVMs \citep[e.g.][]{cortes1995} are a type of classifier that works by 
finding the 
hyperplane in feature space that best separates classes \citep{ball2009}. This hyperplane is 
defined by the vectors to the data points nearest the hyperplane, called the support vectors.
Linear SVMs can be extended to the nonlinear case by using the kernel trick \citep{aizerman1964} to 
transform the features to a 
high-dimensional space so that nonlinear relationships become linear. For this work, we use a 
radial basis function as a kernel, which is a Gaussian function of the Euclidean distance between 
features of different objects.

Probabilities from an SVM can be obtained using Platt scaling \citep{platt1999}, which performs a 
logistic regression of the SVM's decision function scores. Platt scaling is computationally 
intensive, 
requiring an extra cross-validation step to determine the parameters of the transformation 
function. SVMs perform extremely well in general, even in high dimensions, and are highly versatile 
as different kernel functions can be considered to improve feature separation in any particular 
problem. The disadvantage of SVMs lies largely in their computational complexity when computing 
probabilities (if required). \rev{A very recent application of SVMs in astronomy is the 
classification of sources \citep{kurcz2016} for the \emph{WISE} survey \citep{wright2010}.}

\subsubsection{Boosted Decision Trees}
Random forests \citep{breiman2001} and BDTs \citep{friedman2002} are 
ensemble methods built up of a multitude of decision trees \citep[e.g.][]{morgan1963,breiman1984}.

Decision trees construct a model to map input features to output classes using a series of 
decision rules \citep{ball2009}. In our case, the decisions are based on whether or not a given 
feature is in a particular range. The tree is trained by recursively selecting which feature and 
boundary gives the highest information gain in terms of classification. The problem with decision 
trees is that the trees created are often complex and do not generalize well beyond the training 
set, leading to over-fitting. They are also sensitive to small changes in the data, leading to 
large variance in predicted classes.

These problems can be overcome by combining decision trees in an ensemble method. Ensemble learning 
creates a multitude of decision trees on different subsets of data and averages the resulting 
classifications. There are two main approaches of combining classifiers: boosting and bagging.
Random forests are created by bagging, which selects subsets of data with replacement (randomly in 
the case of random forests) on which to perform the 
classification. Boosting repeatedly fits the same dataset but with each 
iteration, it increases the weights of incorrectly classified objects, meaning subsequent 
classifiers focus on difficult cases. We found boosted decision trees (using the AdaBoost algorithm 
\citep{freund1997}) gave the same performance as bagging and were marginally faster in this case.

Probabilities are straightforward to estimate from decision trees: the probability of 
belonging 
to a given class is simply proportional to the fraction of trained objects in that class on the 
corresponding leaf of the tree. With an ensemble method, the probabilities from each of the 
decision trees in the ensemble is averaged to produce a final probability. Boosted decision trees 
are in general excellent classifiers \citep{dietterich2000}. Averaging over a large number of 
classifications makes them 
robust and unbiased estimators. The disadvantage is that they can be computationally intensive, 
although their computation time is still comparable to SVMs and ANNs. \rev{BDTs have been 
successfully applied to the problem of star-galaxy separation in images in \citet{sevilla2015}.}
\\[20pt]

\subsection{Interpreting Machine Learning Results}
\label{sec:metrics}
There are a variety of metrics for measuring the performance of a machine learning algorithm. Every 
machine learning algorithm we consider returns a probability for each classification. It is common 
to 
produce a set of predicted classifications by simply selecting, for each object, the class 
associated with the highest probability. \rev{However, this is not the full picture as the 
probability can be used as a threshold at which an object is considered to belong to a particular 
class. For instance, a very pure sample of type Ia supernovae can be obtained by demanding a 
probability of, say, 95\% before considering the object a type Ia. This purity will always come at 
the cost of completeness, however.} Many metrics common in machine learning literature 
(such as purity or completeness) will depend on this subjective threshold and can be optimized for 
a 
specific class or goal \rev{(for example, obtaining a pure sample of type Ia supernovae for 
cosmology)}. 
Here we use a more general metric, receiver operating characteristic (ROC) curves, to directly 
compare algorithms.

\subsection{ROC Curves}
A confusion matrix is useful for understanding most commonly used machine learning metrics. For a 
binary classification of two classes, positive and negative, the confusion matrix is shown in Table 
\ref{tab:confusion}. An ideal classifier would produce only true positives without any contamination 
from false positives or losses from false negatives. In reality, any classification problem has a 
trade-off between purity and completeness.

\begin{table}[hb]
\caption{Confusion matrix for a classification problem with two classes: positive (P) and negative (N).}
 \label{tab:confusion}
 \centering
 \renewcommand{\arraystretch}{1.6}
 \begin{tabular}{|cc|c|c|}
 \hline
  \multicolumn{1}{|c}{} & \multicolumn{1}{c|}{} & \multicolumn{2}{c|}{True Class}\\
  \multicolumn{1}{|c}{} & \multicolumn{1}{c|}{} & \multicolumn{1}{c}{P} & 
\multicolumn{1}{c|}{N}\\
\hline
  \multirow{2}{*}{Predicted class\negthinspace\negthinspace} & P& True positive (TP) & False positive 
(FP)\\
\cline{3-4}
   & N& False negative (FN) & True negative (TN)\\
 \hline
 \end{tabular}
\end{table}

An excellent method of comparing machine learning algorithms and visualizing this trade-off is to 
use ROC curves \citep{green1966,hanley1982, spackman1989} (see 
\cite{fawcett2004} for an introductory tutorial). These allow one to get an overview of the 
algorithms without having to select an arbitrary probability threshold at which to consider an 
object as belonging to a particular class. ROC curves can only compare one class (assumed to be the 
desired class) against all others, but can easily be constructed for each class in turn. In this paper, all ROC curves are plotted assuming the binary classification of type Ia vs. non-Ia.

ROC curves are constructed by comparing the true positive rate (TPR) against the false positive rate (FPR), as 
the probability threshold is varied. The true positive rate, also called recall, 
sensitivity or completeness, is the ratio between the number of correctly classified 
positive objects and the 
total number of positive objects in the data set,
\begin{equation}
 \rm{TPR}=\frac{TP}{TP+FN}.
\end{equation}

The FPR, also referred to as the false alarm rate or 
$1-$ specificity, is 
the ratio between the number of objects incorrectly classified as positive and the total number of 
negative objects in the data set,
\begin{equation}
 \rm{FPR}=\frac{FP}{FP+TN}.
\end{equation}

It is straightforward to see that a good classification algorithm will maximize the TPR (thus resulting in a more 
complete data set) while simultaneously minimizing the FPR (thus reducing contamination). The area-under-the-curve (AUC) of a ROC curve 
provides a straightforward way to directly compare classifiers. An AUC of 1 represents perfect 
classification, visually represented by a curve which reaches the top left corner. By contrast, a 
diagonal line would correspond to an AUC of 0.5, meaning the classifier does no better than 
random. In the machine learning literature, an AUC higher 
than 0.9 typically indicates an excellent classifier.

In Sec.~\ref{sec:results} we use ROC curves to examine the differences between machine 
learning algorithms and our three different approaches to feature extraction. 

\subsection{Purity and Completeness}
Commonly used metrics for classification include purity and completeness. These are only 
defined for classified objects, meaning some choice of probability threshold must be made. Usually, 
an object's class is selected as that with the highest probability. A more pure subset can be 
obtained by applying a high threshold probability. For any subset chosen, it is useful to determine 
the purity and completeness to evaluate it.

We can use the confusion matrix in Table \ref{tab:confusion} to define purity (also known as 
precision) as
\begin{equation}
 \rm{purity} = \frac{TP}{TP+FP}
\end{equation}
and the completeness, which is equivalent to the TPR, as
\begin{equation}
 \rm{completeness} = \frac{TP}{TP+FN}.
\end{equation}

\subsection{Algorithm Parameter Choices}
In supervised learning, the data are split into a training set, where each object has a known 
class label, and a test set, with unknown objects on which the classifier is run. Choice of 
training set is crucial, as a training set which is not representative of the entire dataset can 
dramatically decrease a classifier's performance (see Sec.~\ref{sec:results}). Additionally, 
almost all machine learning algorithms have hyperparameters that need to be optimized. In this 
work, we use five-fold cross-validation \citep{kohavi1995} to determine the values 
of the hyperparameters (see Table \ref{tab:params} in the Appendix for some example parameter 
values).

\rev{In general, $k$-fold cross-validation divides the training set into $k$ equally sized 
``folds.'' The classifier is trained on $k-1$ folds, leaving the remaining fold as a validation set 
on which the chosen scoring metric (here we use AUC) is evaluated. The score is averaged over all 
$k$ validation sets to produce a final score for the full training set. We use a grid search 
algorithm with cross-validation to find the hyperparameters for each algorithm that maximize AUC.}

\section{Photometric Supernova Classification Pipeline}
For each of the four feature sets (Sec.~\ref{sec:features}) and five machine learning 
algorithms (Sec.~\ref{sec:machine}) considered, we perform the following procedure. 

We split the data into a training and a test set, considering both a representative and 
a non-representative training set separately. The non-representative training set is the same as 
the one used in the SPCC. For a 
representative training set, we randomly select a set of objects the same size as the one in the 
challenge (1103 objects). We were able to repeat the representative case by randomly drawing 
several new training sets and running the full pipeline to see by how much results vary. We did 
not attempt to create a new non-representative training set with the same properties as the 
original.

In the SPCC, photometric host galaxy redshift 
information was provided, so we also investigate the importance of redshift information by running 
the full pipeline with all four feature extraction methods both with and without redshift. This 
is important because accurate photometric redshifts are difficult to obtain and the sensitivity 
of current and proposed classification techniques to the presence of redshift information has not 
yet been established. In the 
case of parametric models 1 and 2 and the wavelet decomposition, the redshift is simply added as an 
additional feature. For the SALT2 case, we fix the redshift in the model to the photometric 
redshift instead of allowing it to vary, thus better constraining the other parameters for type 
Ia's. 

Before running any machine learning algorithm, we rescale the features with a standard 
scaling. We scale the features by removing the mean (centering on zero) and scaling to unit 
variance (dividing by the standard deviation). After rescaling, we run each of the machine learning 
algorithms outlined in Sec.~\ref{sec:machine} using cross-validation to determine the optimal 
hyperparameters. We then compare the results using ROC curves.

\section{Results}
\label{sec:results}

\begin{figure*}
\centering
\subfigure[SALT2 model, no redshift]{
 \label{fig:templates_rep_no_z}
 \includegraphics[width=0.45\textwidth]{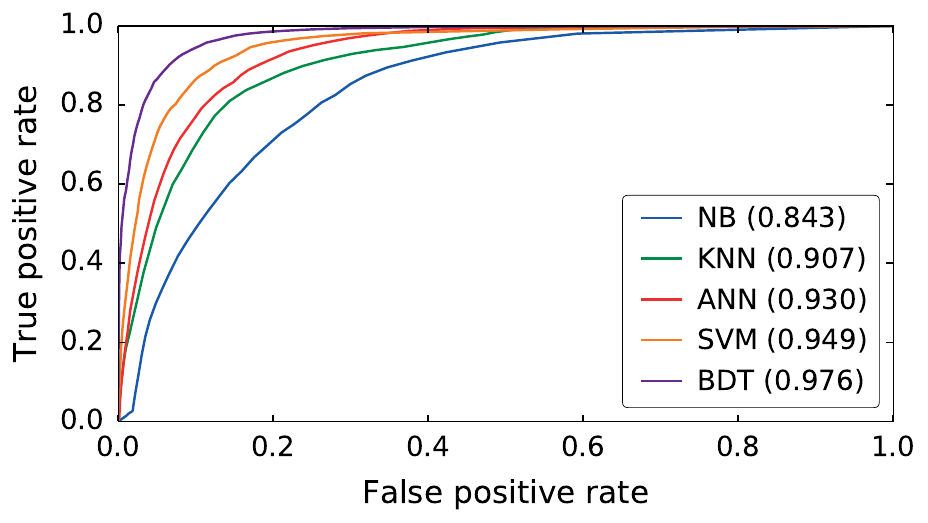}}
\subfigure[SALT2 model, with redshift]{
 \label{fig:templates_rep}
 \includegraphics[width=0.45\textwidth]{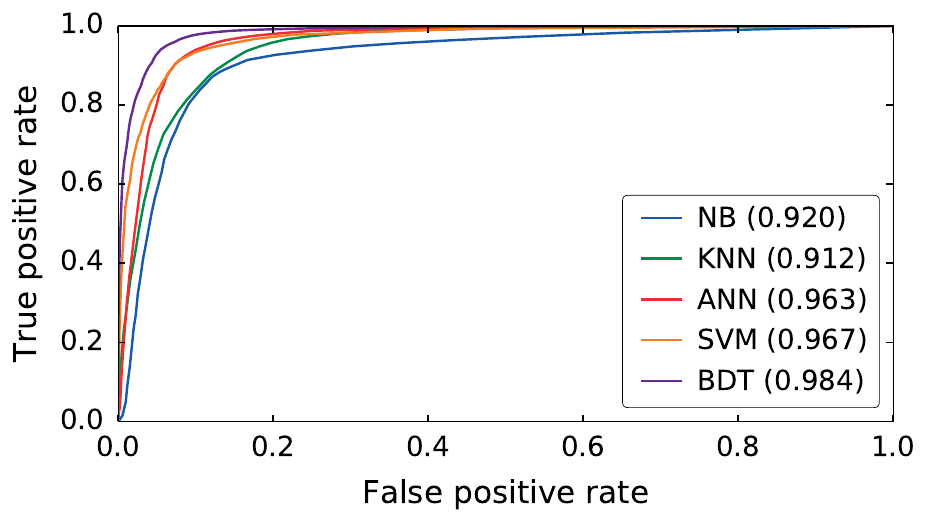}}
 
  \subfigure[Parametric model 1, no redshift]{
 \label{fig:newling_rep_no_z}
 \includegraphics[width=0.45\textwidth]{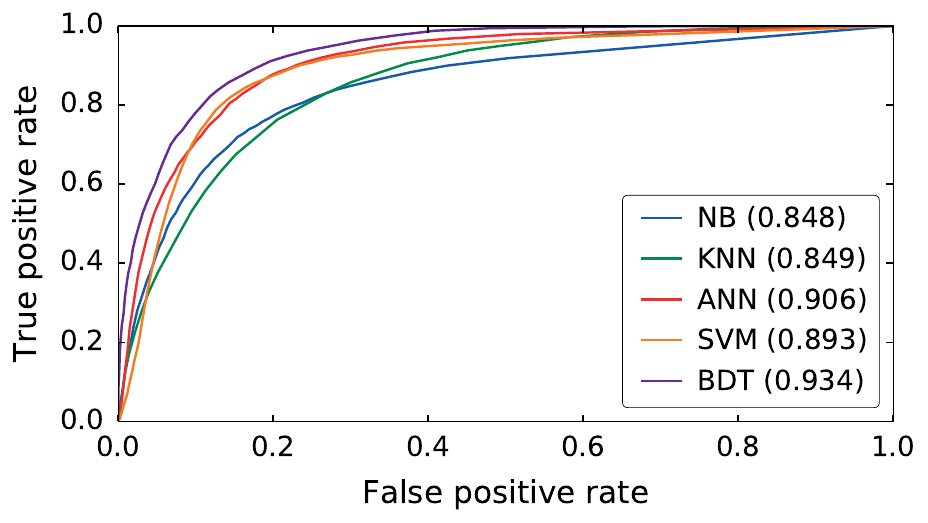}}
 \subfigure[Parametric model 1, with redshift]{
 \label{fig:newling_rep}
 \includegraphics[width=0.45\textwidth]{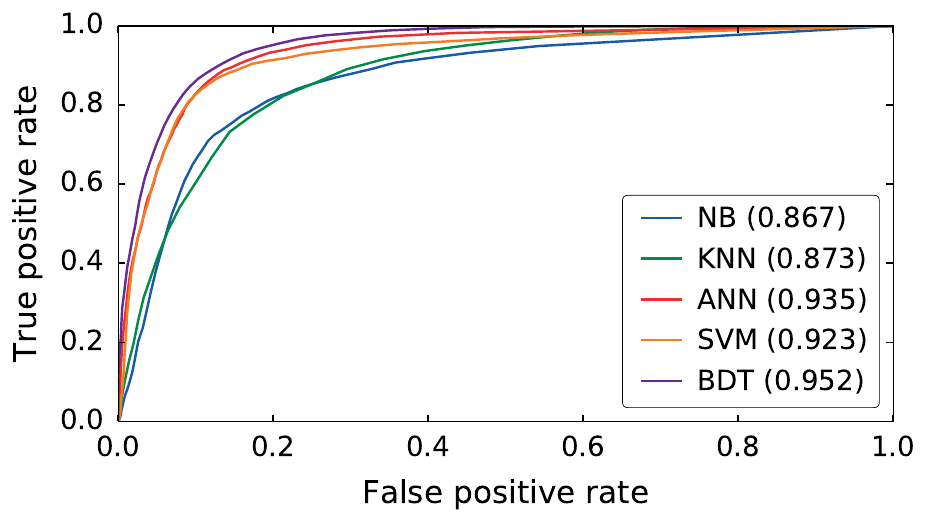}}
 
  \subfigure[Parametric model 2, no redshift]{
 \label{fig:karpenka_rep_no_z}
 \includegraphics[width=0.45\textwidth]{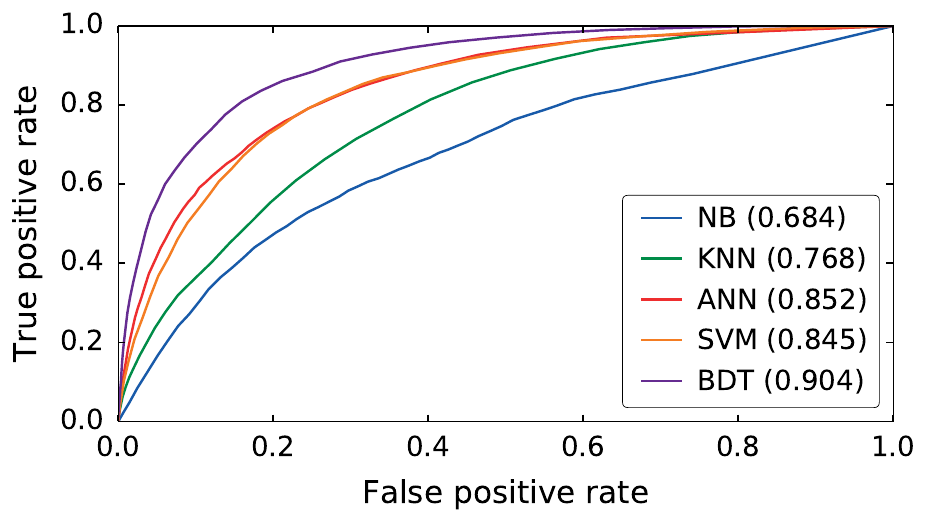}}
 \subfigure[Parametric model 2, with redshift]{
 \label{fig:karpenka_rep}
 \includegraphics[width=0.45\textwidth]{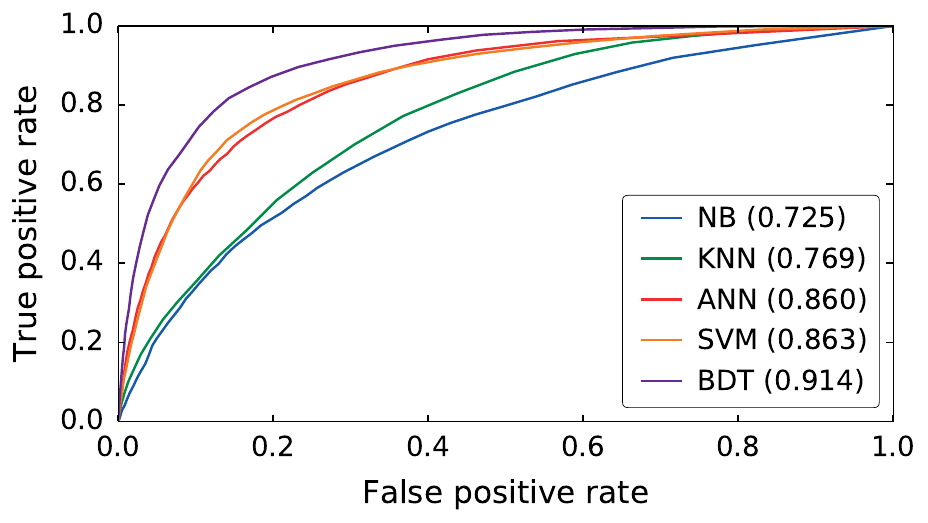}}

 \subfigure[Wavelets, no redshift]{
 \label{fig:wavelets_rep_no_z}
 \includegraphics[width=0.45\textwidth]{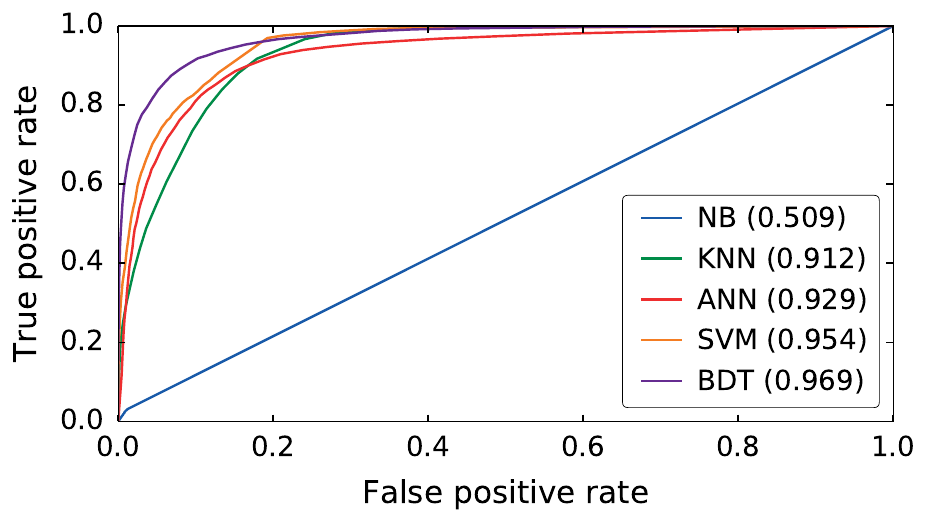}}
\subfigure[Wavelets, with redshift]{
 \label{fig:wavelets_rep}
 \includegraphics[width=0.45\textwidth]{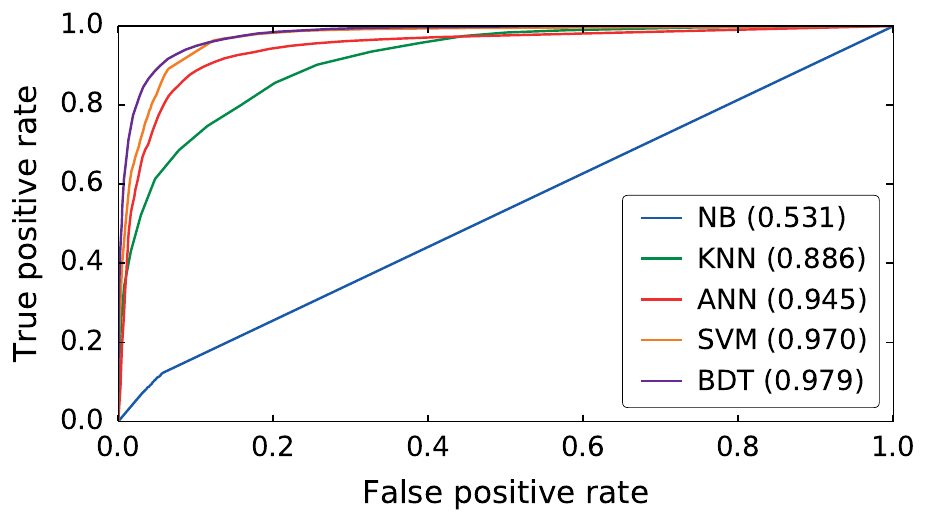}}

\caption{Receiver operating characteristic (ROC) curves for the various feature extraction 
methods, 
using a representative training set without (first column) and with (second column) redshift data 
included. Each curve represents a 
different machine learning algorithm (as outlined in Sec.~\ref{sec:machine}) with the 
area-under-curve (AUC) score in brackets. An AUC of 1 indicates perfect classification and 
corresponds to a curve close to the top left corner. It is clear that boosted decision trees 
(BDTs), 
support vector machines (SVMs) and artificial neural networks (ANNs) are the best-performing 
algorithms, superior to \emph{k}-nearest neighbors (KNN) and naive Bayes (NB). The SALT2 and 
wavelet 
feature sets provide the best classification.}
\label{fig:roc_z}
\end{figure*}

Figure \ref{fig:roc_z} shows the ROC curves for a representative training set, both without and 
with redshift 
information included. From the 
ROC curves, it is clear that not all machine learning algorithms perform equally and indeed, in 
every case the BDTs outperform the other algorithms. These ROC curves also 
illustrate that if the fundamental assumptions of an algorithm are broken, it performs poorly. 
Naive 
Bayes performs reasonably well in most cases, but is barely better than random in the case of the 
wavelets, which are highly non-Gaussian in their distribution. We can also see that the SALT2 
features and the wavelet features outperform the parametric models, implying that it is best to 
use either a highly model-dependent approach, making use of prior knowledge, 
or with a highly model-independent approach for supernova classification.

\begin{figure*}
 \centering
\subfigure[All types]{
 \label{fig:venn_all}
 \includegraphics[width=0.45\textwidth]{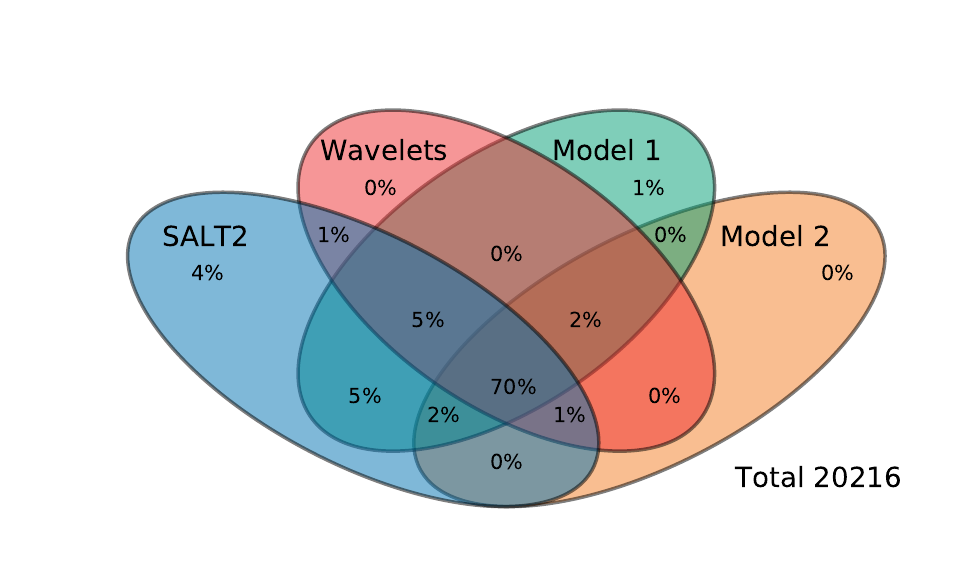}}
 \subfigure[Type Ia]{
 \label{fig:venn_Ia}
 \includegraphics[width=0.45\textwidth]{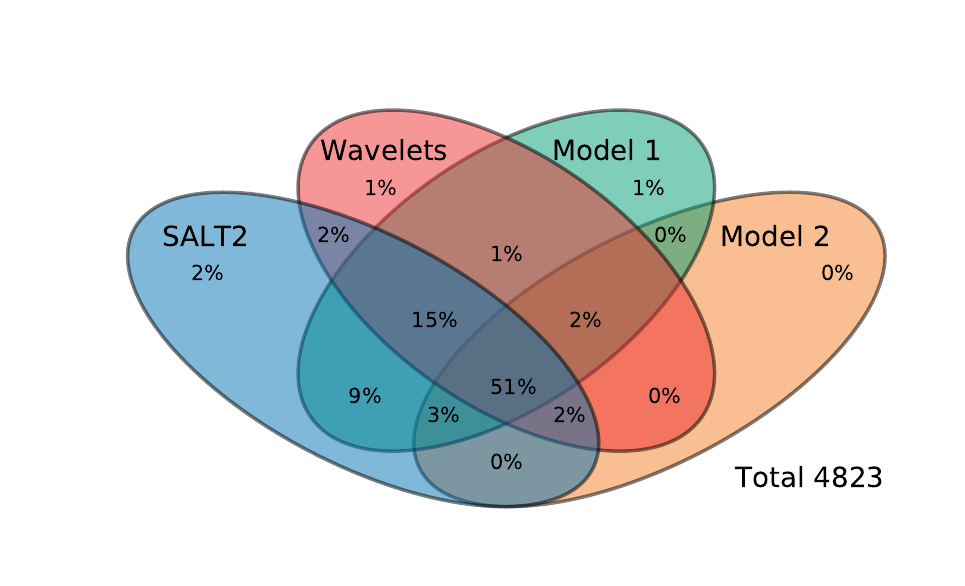}}
 \subfigure[Type II]{
 \label{fig:venn_II}
 \includegraphics[width=0.45\textwidth]{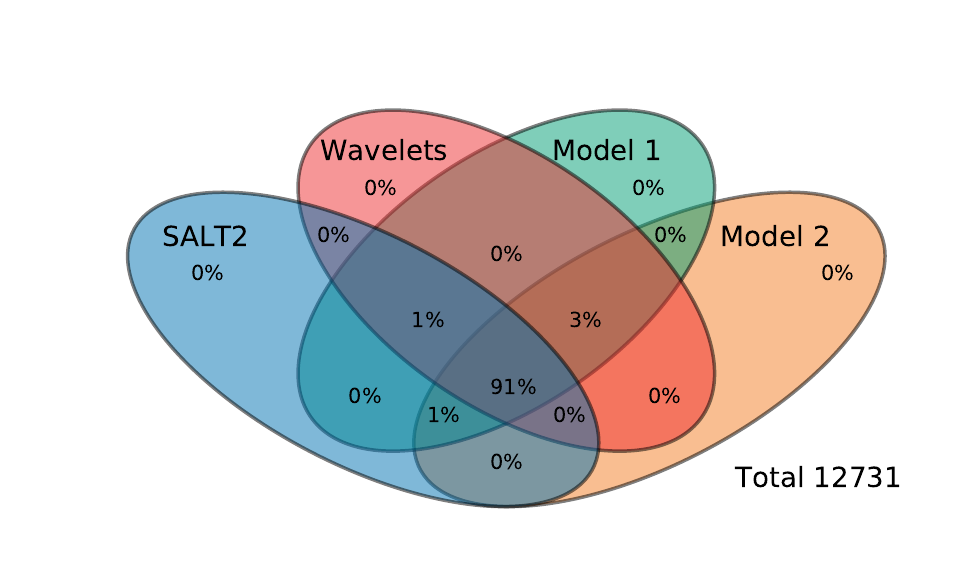}}
\subfigure[Type Ibc]{
 \label{fig:venn_Ibc}
 \includegraphics[width=0.45\textwidth]{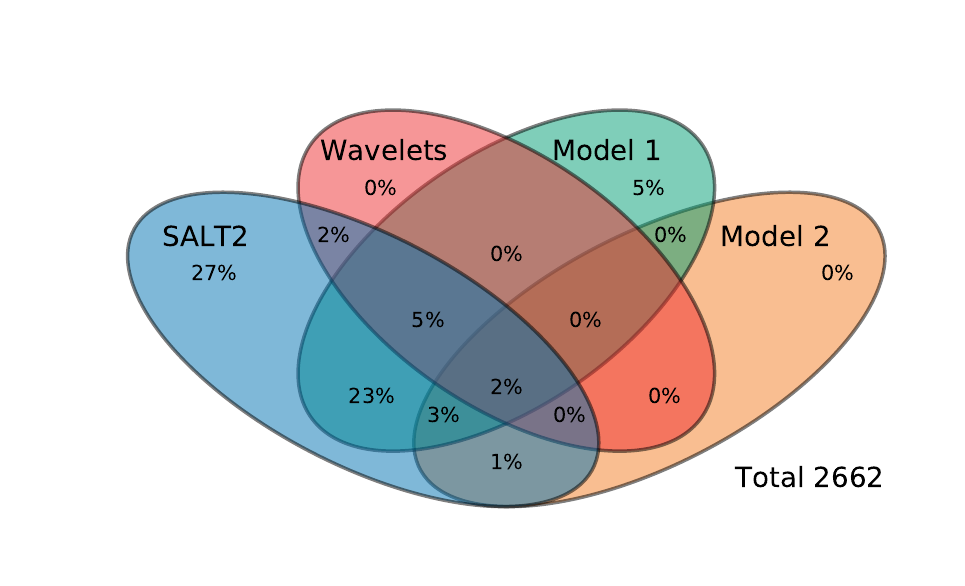}}

\caption{Venn diagrams showing the number of correctly classified objects for each 
feature 
extraction method using boosted decision trees, with each area indicating the percentage of 
objects correctly classified. This is repeated for all objects and for each type of supernova 
individually. It is clear that, for example, type II supernovae are easy to classify, with all 
feature extraction methods agreeing on the types and achieving excellent accuracy, while type Ibc's 
are much more difficult to classify.} 
\label{fig:venn}
\end{figure*}

Figure \ref{fig:venn} shows the Venn diagram for object classification for each of the 
four feature extraction methods, using BDTs. Objects were classified according to 
the class with highest probability. From this, one can see that type II supernovae are 
straightforward to classify, with all feature extraction methods achieving excellent accuracy. Type 
Ia's are somewhat more difficult, although the SALT2, Model 1 and wavelets methods agree well. 
There is little agreement between classifiers when it comes to the type Ibc's. This is 
unsurprising as Ibc light curves are often similar to Ia's and there are relatively few Ibc's in 
the dataset. The SALT2 method does significantly better than the others at identifying Ibc's.

As stated in Sec.~\ref{sec:metrics}, the probability threshold at which you select an object's 
class is 
arbitrary, meaning that the given class for each supernova will change depending on the 
threshold used. It is well known that commonly used metrics such as accuracy, purity and 
completeness are dependent on the threshold used and do not give a complete picture (hence why we 
advocate the use of the AUC as a comparison metric).
However as an example of the values of these metrics for the SPCC dataset, we can vary the 
threshold probability and insist on a 90\% pure type Ia dataset. Then the corresponding 
completeness 
is 85\%, 41\%, 7.8\%, 83\% for SALT 2, Model 1, Model 2 and wavelets respectively. However, if a 
more complete dataset is more important, we find that demanding a 90\% completeness results in a 
purity of 87\%, 77\%, 75\%, 87\% respectively.

\subsection{The Effect of Redshift}
\begin{figure*}[hb!]
\centering
 \includegraphics[width=0.8\textwidth]{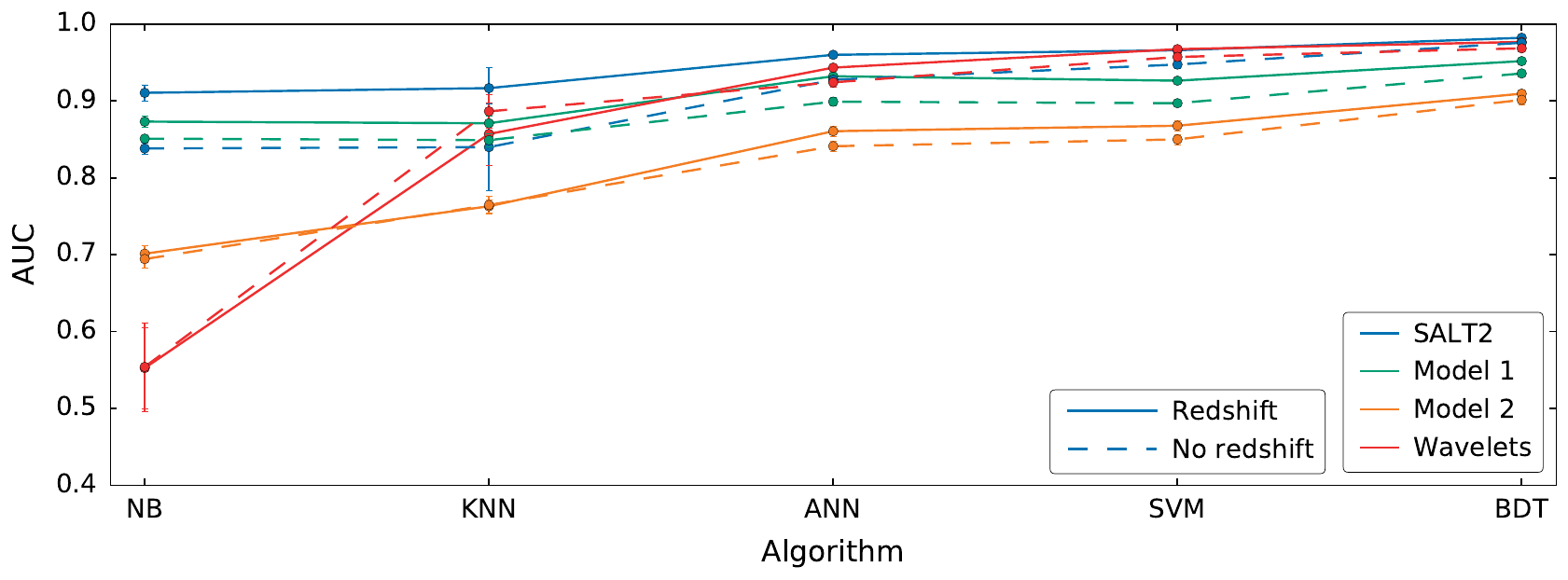}
 \caption{Area-under-curve (AUC) as a function of feature set and machine learning algorithm. 
Higher AUC scores correspond to better classification. Solid lines indicate redshift was included as a
feature, while dashed lines represent feature sets without redshift information. The five machine 
learning algorithms considered are on the $x$-axis, while each color corresponds to a different 
feature extraction method. Error bars are from the standard deviation of the AUC from ten runs with 
randomized training sets. When using BDT, the SALT2 and wavelet features are able 
to classify equally well with or without redshift. By comparison, the SALT2 features are highly 
sensitive to redshift when using, for example, the NB or KNN algorithm.}
\label{fig:auc_redshift}
\end{figure*}

The effect of redshift is 
summarized succinctly in Fig.~\ref{fig:auc_redshift}, comparing all feature extraction 
methods and machine learning algorithms. Figure \ref{fig:auc_redshift} shows that providing 
redshift information plays an important role 
for SALT2 features, mildly improves the results of the parametric features, and is fairly 
unimportant to the wavelet features. The notable exception to this is when considering the best 
machine learning algorithm, BDTs. With this algorithm, including redshift 
information is not essential as the algorithm is capable of differentiating between classes 
without redshift information.

This raises the interesting point that with an algorithm with poor performance, such as KNN, 
redshift information is crucial to classify well with the SALT2 model. However, a better 
machine learning algorithm eliminates the need for redshift information (under the assumption of a 
representative training set). This is important since obtaining reliable redshifts for every object 
in future surveys will be challenging. We have shown that it will be possible to obtain a 
relatively pure subsample of supernovae without redshift information, on which to focus follow-up 
observations.

It is also interesting to note that the wavelet method, which uses no prior knowledge of 
supernovae, is able to perform as well as the current state-of-the-art classification technique, 
based on supernova templates. This is promising since the simulated dataset used here was 
constructed using largely the same known templates, so it is expected that the SALT2 model should perform 
well. However, the wavelet approach requires no previous knowledge and so should perform well with 
new datasets. This also suggests wavelet approaches to 
classification are likely to be useful for broader transient classification, as also noted in 
\cite{varughese2015}.

\begin{figure*}[ht!]
\centering
 \includegraphics[width=0.8\textwidth]{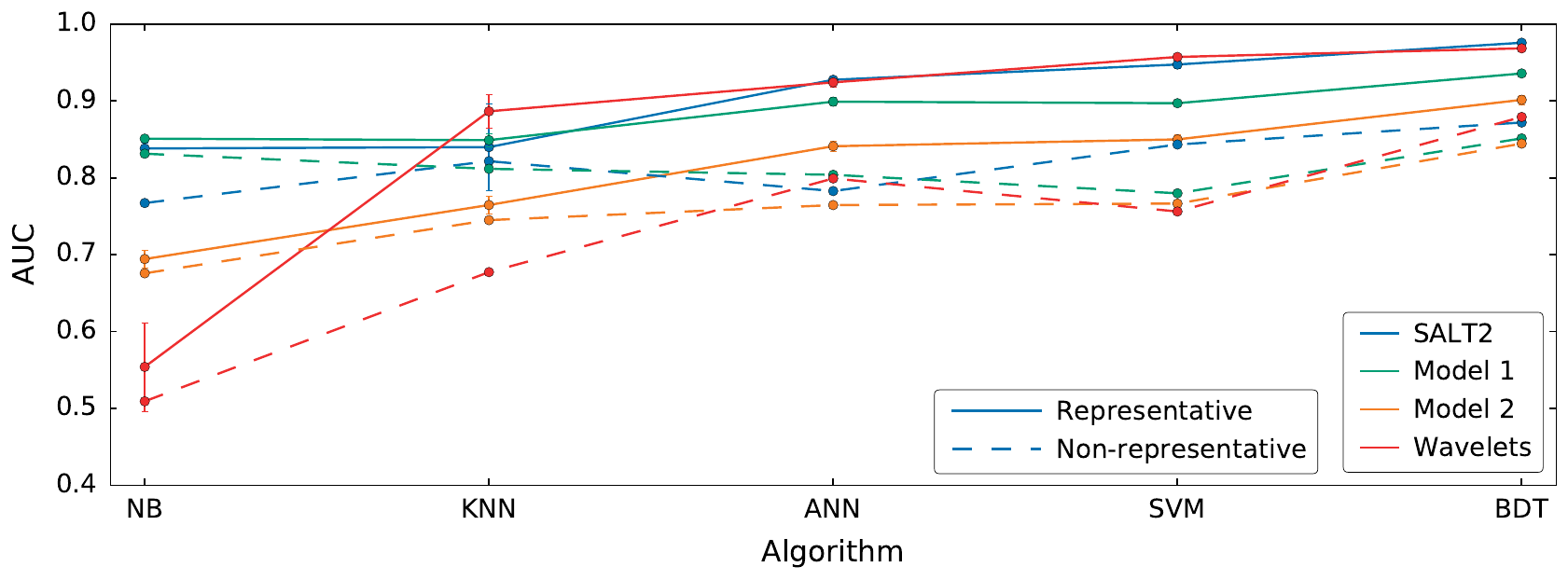}
 \caption{Area-under-curve (AUC) as a function of feature set and machine learning algorithm. 
Higher AUC scores correspond to better classification. Solid lines indicate 
a representative training set was used, while dashed lines indicate the use of the 
non-representative training set of \cite{kessler2010a,kessler2010b}. The five machine 
learning algorithms considered are on the $x$-axis, while each color corresponds to a different 
feature extraction method. Error bars for the representative training set case are the standard 
deviation from ten runs with randomized training sets. It is clear that any feature extraction 
method or machine learning algorithm used is highly sensitive to non-representativeness in the 
training set.}
\label{fig:auc_repr}
\end{figure*}

\subsection{The Effect of Representativeness}
We also investigate the effect of using a representative or non-representative training set. It is 
well known that most machine learning algorithms underperform when the training set is in 
any way not representative of the distribution of features. The training set from the SPCC was known 
to be biased in the way spectroscopic follow-up 
datasets usually are, in that only the brightest objects tend to be followed up due to telescope 
time constraints and the difficulty in obtaining good spectra. It is not surprising then that, as 
shown in Fig.~\ref{fig:auc_repr}, all feature extraction methods and all algorithms perform 
significantly worse with the non-representative training set. It should be noted that redshift 
information was not
provided here, which only mitigates the effect of non-representativeness in the case of the SALT2 
model. This is because the SALT2 model is already based on a large training set of supernovae and 
thus, if redshift information is given, will produce similar sets of features.

The clear conclusion is that the more representative the training set, the better any 
classification method will perform. As the training set was only around 5\% of the size of the full 
dataset, we would argue that if spectroscopic follow-up time is limited, it is rather better to 
obtain fewer, fainter objects of all types, than to get a larger sample of brighter objects (as 
noted in \cite{richards2012}).

\rev{However, the effect of non-representativeness, also called the ``domain adaptation problem'' 
or ``sample selection bias,'' can be mitigated using modern techniques in transductive 
transfer learning \citep{pan2010}. These include reweighting techniques to reweight the 
data in the 
test set to more closely match that of the training set and techniques to learn new sets of 
features that minimize the difference between the two datasets. Transfer learning has been 
successfully used in astronomy in \citet{sasdelli2015}. An in-depth study of transfer 
learning is beyond the scope of this paper but will likely prove useful for dealing with 
non-representativeness in the training sets of future supernova surveys.}

\subsection{\rev{Relative Feature Importance}}
\begin{figure*}
\centering
\subfigure[SALT2]{
 \label{fig:importances_templates}
 \centering
 \includegraphics[width=0.8\textwidth]{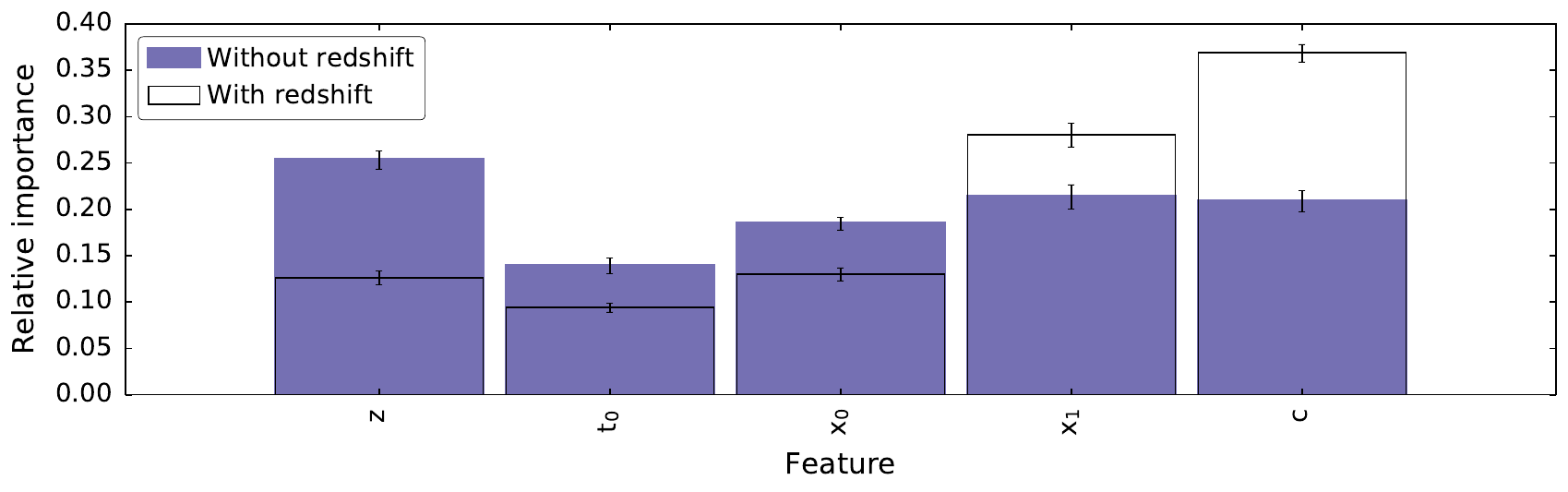}}
\subfigure[Parametric model 1]{
 \label{fig:importances_newling}
 \centering
 \includegraphics[width=0.8\textwidth]{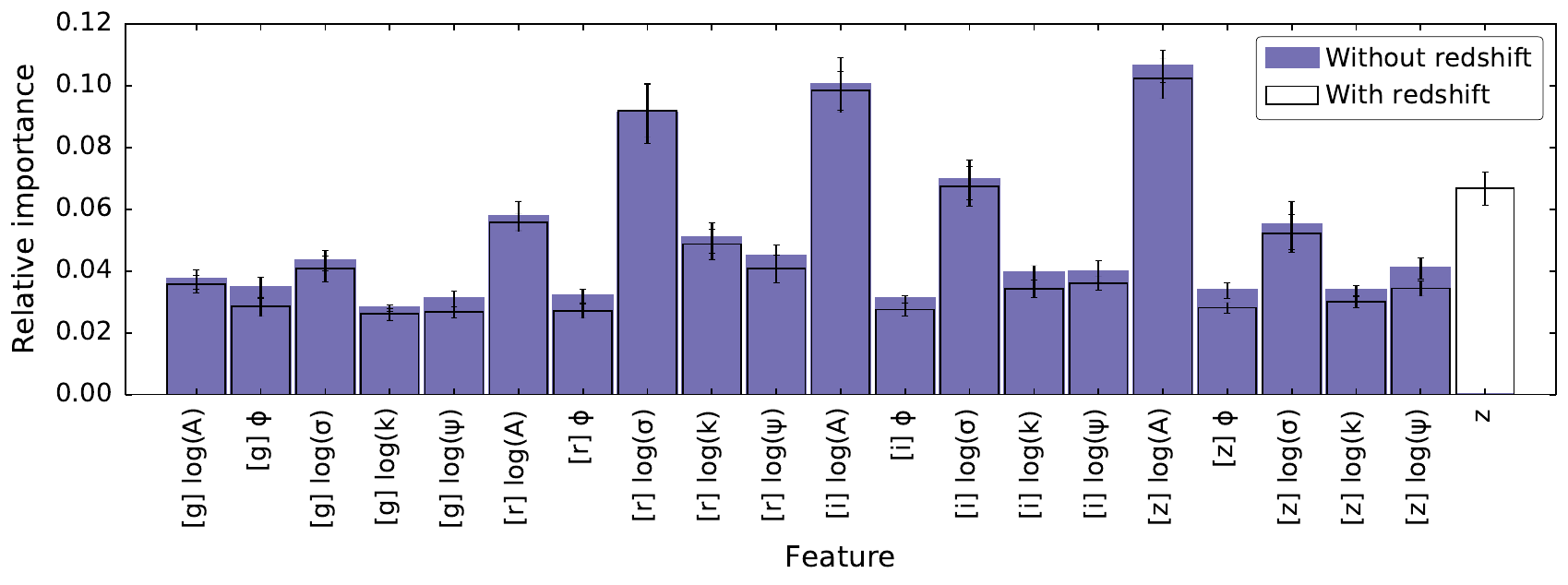}}
\subfigure[Parametric model 2]{
 \label{fig:importances_karpenka}
 \centering
 \includegraphics[width=0.8\textwidth]{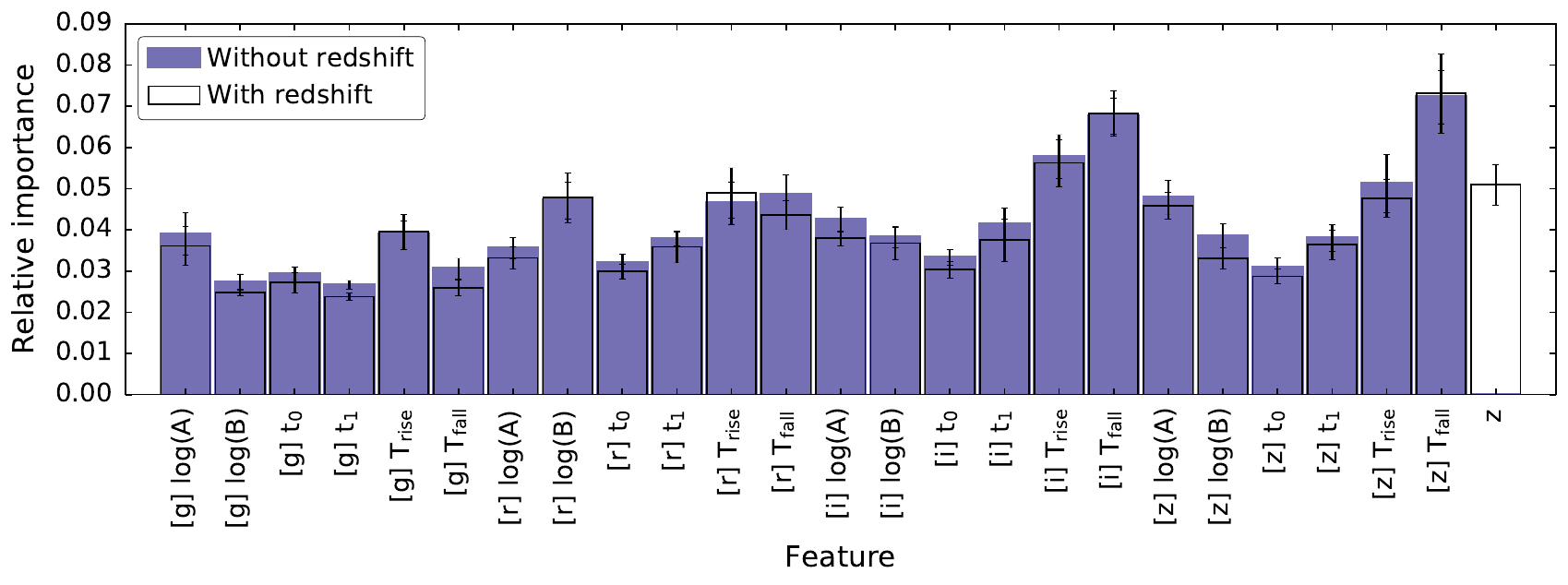}}
\subfigure[Wavelets]{
 \label{fig:importances_wavelets}
 \centering
 \includegraphics[width=0.8\textwidth]{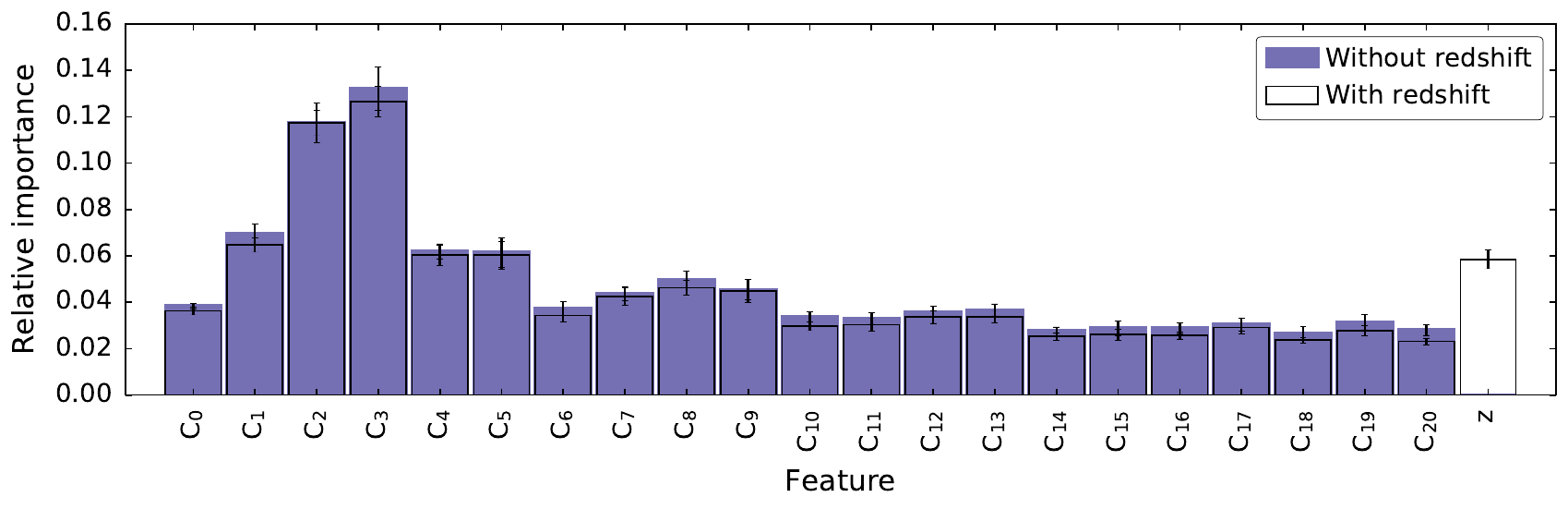}}

\caption{Relative importances of individual features with and without redshift information, as 
obtained from the boosted decision trees algorithm. The SALT2 model is able to estimate the 
redshift, which is thus still used as a feature even when no redshift information is provided (and 
a broad, flat prior is used for the fitting). For parametric models 1 and 2, the filter is given in 
square brackets before the parameter. For the wavelets, the 20 principal components used are 
labeled $C_1$ to $C_{20}$.}
\label{fig:importances}
\end{figure*}

It is possible to investigate the relative importance of each feature in a feature set 
using BDTs \rev{using the ``gini importance'' (also known as ``mean decrease 
impurity''), described in \cite{breiman1984}.} This is illustrated for 
each feature set in 
Fig.~\ref{fig:importances}, both with and without redshift. \rev{While there are many other methods 
to determine the 
relevance of each feature (see \cite{li2016} for a summary), we only use 
this simple measure of feature relevance to gain insight about the features and use the full 
classification pipeline to evaluate the effectiveness of our chosen features, rather than 
performing initial feature selection based on relevance.}

For the SALT2 model, the most important features are the shape and color parameters. Notice that 
counterintuitively, the importance of redshift goes \emph{down} when external redshift information 
is added. By adding redshift information, the Ia's in the dataset are better fit by the SALT2 
model as expected, resulting in accurate estimates of the stretch and color parameters. On the 
other hand, with the redshift fixed the best fits for the non-Ia's result in unphysical stretch 
and color parameters. Because the distributions of these important 
features now differ much more between the non-Ia's and Ia's, they increase in importance when 
classifying. Because all the importances must add up to one, this necessarily dictates a decrease 
in the importance of other features, including the redshift.

There are a few features which are more important for the parametric models, but none notable. 
Interestingly, it is the third and fourth principal components for the wavelet features which are 
most important, while all others have fairly equivalent importance.

\section{Conclusions}
\label{sec:conclusions}
Classification of photometric supernovae is an important problem in light of current and upcoming 
surveys. We have compared several different approaches to this problem, some based on current 
techniques and some new. We have shown that in terms of feature extraction, both SALT2 model fits 
and 
wavelet decomposition outperform parametric fits. Out of the five representative machine learning 
algorithms we used, boosted decision trees performed the best, followed by support vector 
machines and artificial neural networks. The SALT2 fits and the wavelet approach both achieved an 
AUC of 0.98, representing excellent classification. The SALT2 and wavelet methods can both produce 
a dataset that is about 90\% 
pure and 84\% complete, although the purity can be increased at the cost of completeness. 

Importantly, we found that with a powerful ensemble 
algorithm like boosted decision trees and a representative training set, redshift information does 
not improve classification performance, meaning a dataset can be well classified without any 
redshift information at all. This is extremely useful for current and upcoming surveys where 
reliable host galaxy photometric redshifts may be difficult to obtain. 

Although redshift 
information is needed for cosmology, this means a well-classified subset of supernovae will be 
available without any redshift information, with which follow-up studies can be done if necessary. 
This also implies uncertainties in classification will be completely uncorrelated with 
uncertainties in redshift estimates.
It also means there will be a large sample of robustly classified supernovae available for core 
collapse 
research.

On the other hand, with all algorithms and feature extraction methods considered, a 
representative training set is crucial for good performance. The training set used was very small 
(around 5\% of the dataset) suggesting that a spectroscopic follow-up program should rather focus 
on observing fainter objects of all types rather than obtaining a larger training set 
which is not representative. 

There are many further improvements one can consider in both the feature extraction methods and the 
machine learning algorithms. We have provided a framework here for directly and easily comparing 
these and any future approaches to the photometric supernova classification problem.

In future work we will apply this pipeline to SDSS data and compare against current 
photometric classifications. Additionally, the pipeline developed here will be crucial in 
investigating how changes in observing strategy affect classification for LSST.

\acknowledgments
\section*{Acknowledgments}
We thank Michael Betancourt, Rahul Biswas, Saurabh Jha, Bob Nichol, Stephen Smartt and Mark Sullivan 
for useful discussions. We acknowledge support from STFC for UK participation in LSST through grant 
ST/N00258X/1 and travel support provided by STFC for UK participation in LSST through grants 
ST/L00660X/1 and ST/M00015X/1. HVP and ML were supported by the European Research Council under the 
European Community's Seventh Framework Programme (FP7/2007-2013)/ERC grant agreement no 
306478-CosmicDawn. OL and ML acknowledge a European Research Council Advanced Grant FP7/291329 
(TESTDE) and support from STFC.

\bibliographystyle{apj}
\bibliography{main}

\appendix
\section{Hyperparameter Selection}

Table \ref{tab:params} shows the values of the hyperparameters for each machine 
learning algorithm for an example case. These hyperparameters are selected using cross-validation 
(as described in Sec.~\ref{sec:machine}) to maximize the AUC. These values vary slightly depending 
on training set used.
\begin{table}[hb!]
 \centering
 \caption{Hyperparameters for each machine learning algorithm and each feature set for the case 
where redshift is included and a representative dataset is used.}
\label{tab:params}
 \begin{tabular}{c|cccc}
 \hline
  &SALT2&Model 1&Model 2&Wavelets\\
  \hline
  NB& ... & ... & ... & ... \\
  KNN& n\_neighbors=141& n\_neighbors=101& n\_neighbors=46 & n\_neighbors=121\\
  SVM& C=0.56, $\gamma$=1.78& C=1780,$\gamma$=0.0032& C=31.6,$\gamma$=0.0032 & 
C=0.56,$\gamma$=1.78\\
  ANN& neurons=110 & neurons=110 & neurons=115 & neurons=80\\
  BDT& n\_estimators=40 & n\_estimators=75 & n\_estimators=65 & n\_estimators=70\\
  \hline
 \end{tabular}

\end{table}

\end{document}